\documentclass[lettersize,journal]{IEEEtran}
\usepackage{hyperref}
\usepackage{amsmath,amsfonts}
\usepackage{algorithmic}
\usepackage{algorithm}
\usepackage{array}
\usepackage[caption=false,font=normalsize,labelfont=sf,textfont=sf]{subfig}
\usepackage{textcomp}
\usepackage{stfloats}
\usepackage{url}
\usepackage{verbatim}
\usepackage{graphicx}
\usepackage[table]{xcolor}
\usepackage{amssymb}
\usepackage{multirow, threeparttable, booktabs, makecell}
\usepackage{color}
\usepackage{enumitem}
\usepackage{placeins}
\usepackage{afterpage}
\usepackage[numbers,sort&compress]{natbib}
\usepackage{textcomp}
\usepackage[normalem]{ulem}
\usepackage{adjustbox}
\usepackage[switch]{lineno}
\IEEEoverridecommandlockouts

\begin{document}
% \linenumbers
% \switchlinenumbers    % 支持双栏行号

% \title{MF-AED-AEC: SPEECH EMOTION RECOGNITION BY LEVERAGING MULTIMODAL FUSION, ASR ERROR DETECTION, AND ASR ERROR CORRECTION}
\title{M\textsuperscript{4}SER: Multimodal, Multirepresentation, Multitask, and Multistrategy Learning \\
for Speech Emotion Recognition}

\author{Jiajun He, Xiaohan Shi, Cheng-Hung Hu, Jinyi Mi, Xingfeng Li and Tomoki Toda,~\IEEEmembership{Senior Member, IEEE}
        % <-this % stops a space
\thanks{This work was supported in part by JST CREST Grant Number JPMJCR22D1, Japan, and JSPS KAKENHI Grant Number 21H05054.}% <-this % stops a space
\thanks{Jiajun He, Xiaohan Shi,  Cheng-hung Hu, and Jinyi Mi are with the Graduate School of Informatics, Nagoya University, Nagoya 464-8601, Japan (e-mail: \{jiajun.he, xiaohan.shi, hu.chenghung, mi.jinyi\}@g.sp.m.is.nagoya-u.ac.jp).}
\thanks{Xingfeng Li is with the Faculty of Data Science, City University of Macau, Macau 999078, China (e-mail: xfli@cityu.edu.mo).}
\thanks{Tomoki Toda is with the Information Technology Center, Nagoya University, Nagoya 464-8601, Japan (e-mail: tomoki@icts.nagoya-u.ac.jp).}
}

\markboth{Journal of \LaTeX\ Class Files,~Vol.~14, No.~8, August~2021}%
{Shell \MakeLowercase{\textit{et al.}}: A Sample Article Using IEEEtran.cls for IEEE Journals}


\maketitle
% \renewcommand{\arraystretch}{1.5} % Adjust the factor as needed

% ==========abstract and keywords==========
\begin{abstract}
Multimodal speech emotion recognition (SER) has emerged as pivotal for improving human\textendash machine interaction. Researchers are increasingly leveraging both speech and textual information obtained through automatic speech recognition (ASR) to comprehensively recognize emotional states from speakers. Although this approach reduces reliance on human-annotated text data, ASR errors possibly degrade emotion recognition performance. 
To address this challenge, in our previous work, we introduced two auxiliary tasks, namely, ASR error detection and ASR error correction, and we proposed a novel multimodal fusion (MF) method for learning modality-specific and modality-invariant representations across different modalities. 
Building on this foundation, in this paper, we introduce two additional training strategies. First, we propose an adversarial network to enhance the diversity of modality-specific representations. Second, we introduce a label-based contrastive learning strategy to better capture emotional features.
We refer to our proposed method as M$^4$SER and validate its superiority over state-of-the-art methods through extensive experiments using IEMOCAP and MELD datasets.
\end{abstract}

% ==========abstract and keywords==========

% ==========Introduction==========
\vspace{-5mm}
\section{Introduction}

% ================================
% 新的introduction
% 情感识别很重要：讲一些场景和应用OK。

% 多模态语音情感识别（SER）作为智能人机交互系统的核心，在医疗保健和智能客户服务等领域中具有重要作用。通过分析语音、语言、面部表情和姿势等生理信号，多模态SER能够识别和理解人类的情感状态，因此成为当前研究人员广泛关注的关键技术之一。本文将重点研究语音和文本这两种最常见的模态，用于实现多模态语音情感识别。

\IEEEPARstart{M}{ultimodal} speech emotion recognition (SER) serves as a cornerstone in intelligent human\textendash computer interaction systems, playing a crucial role in fields such as healthcare and intelligent customer service \cite{ geetha2024multimodal, de2023ongoing}. By analyzing physiological signals such as speech, language, facial expressions, and gestures, multimodal SER can identify and understand human emotional states, making it a key technology of widespread interest among researchers nowadays.

% =======pipeline=======================
\begin{figure}[t]
  \centering
  \includegraphics[width=1\columnwidth]{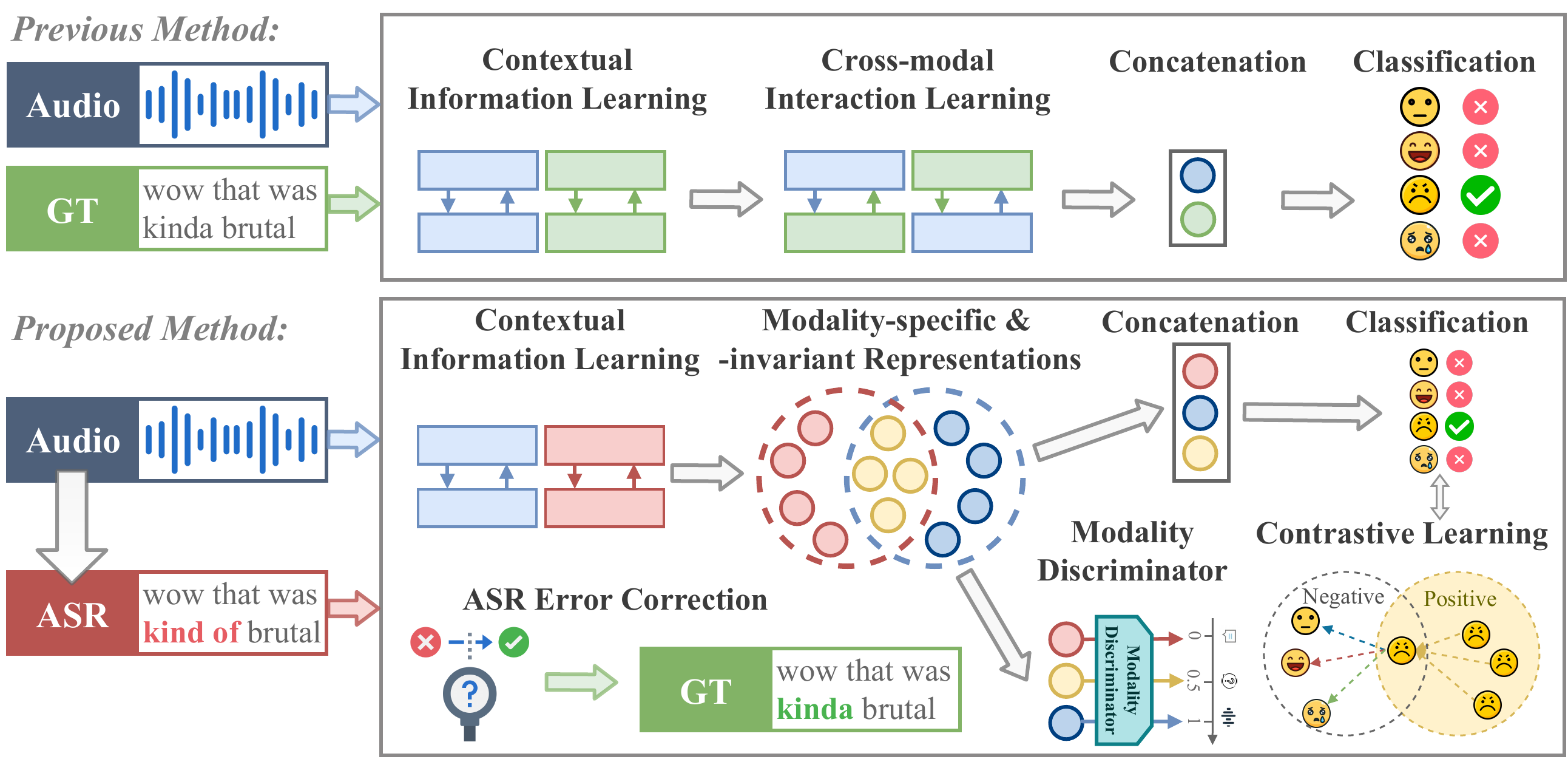}
  \vspace{-6mm}
  \caption{Illustration of the difference between previous multimodal SER methods and our proposed method. ``ASR" and ``GT"  denote ASR and ground truth transcripts, respectively.}
  \vspace{-5mm}
  \label{fig:pipeline}
\end{figure}
% =======pipeline=======================

With the rapid advancement of deep learning, SER has evolved into end-to-end (E2E) systems capable of directly processing speech signals and outputting emotion classification results. 
In recent years, the success of self-supervised learning (SSL) has led to the development of a series of speech-based pretrained models such as wav2vec \cite{schneider2019wav2vec}, HuBERT \cite{hsu2021hubert}, and WavLM \cite{chen2022wavlm}, achieving the current state-of-the-art (SOTA) performance in SER tasks.

Recent research has shown that single-modal approaches struggle to meet the increasing demands for SER owing to their inherent complexity. As a solution, multimodal information has emerged as a viable option because it offers diverse emotional cues. A common strategy involves integrating speech and text modalities to achieve multimodal SER. Text data can be obtained through manual transcription or automatic speech recognition (ASR), which have been widely adopted in recent studies to reduce reliance on extensive annotated datasets.
Although text-based pretrained models such as BERT \cite{devlin2018bert} and RoBERTa \cite{liu1907roberta} have excelled in multimodal SER tasks, the accuracy of ASR remains equally crucial for achieving precise emotion recognition (ER) results. To mitigate the impact of ASR errors, Santoso et al. \cite{santoso2022speech} have integrated self-attention mechanisms and word-level confidence measurements, particularly reducing the importance of words with high error probabilities, thereby enhancing SER performance. However, this approach heavily depends on the performance of the ASR system, thus limiting its generalizability. Lin and Wang \cite{lin2023robust} have proposed an auxiliary ASR error detection task aimed at determining the probabilities of erroneous words to adjust the trust levels assigned to each word in the ASR hypothesis, thereby improving multimodal SER performance. However, this method focuses solely on error detection within the ASR hypothesis without directly correcting these errors, indicating that it does not fully enhance the coherence of semantic information.

On the other hand, how to effectively integrate speech and text modalities has become a key focus. Previous methods have included simple feature concatenation, recurrent neural networks (RNNs), convolutional neural networks (CNNs), and cross-modal attention mechanisms \cite{geetha2024multimodal}. Although these methods have shown some effectiveness, they often face challenges arising from differences in representations across different modalities.
In recent multimodal tasks such as sentiment analysis \cite{hazarika2020misa, yang2022disentangled}, researchers have proposed learning two distinct types of representation to enhance multimodal learning. 
The first is a representation tailored to each modality, that is, a modality-specific representation (MSR).
The second is modality-invariant representation (MIR), aimed at mapping all modalities of the same utterance into a shared subspace. This representation captures commonalities in multimodal signals as well as the speaker's shared motives, which collectively affect the emotional state of the utterance through distribution alignment. By combining these two types of representation, one can obtain a comprehensive view of multimodal data can be provided for downstream tasks \cite{yang2022learning}. 
However, these methods focus on utterance-level representations, which, while effective for short and isolated utterances, often fail to capture fine-grained emotional dynamics, especially in long and complex interactions. Such coarse-grained representations fail to capture subtle temporal cues and modality-specific variations that are crucial for accurate emotion recognition.
Moreover, mapping entire utterances to a shared or modality-specific space using similarity loss oversimplifies the underlying multimodal relationships and hinders effective cross-modal alignment.

On the basis of the above observations, in our previous work \cite{10446548}, we introduced two auxiliary tasks: ASR error detection (AED) and ASR error correction (AEC). Specifically, we introduced an AED module to identify the positions of ASR errors. Subsequently, we employed an AEC module to correct these errors, enhancing the semantic coherence of ASR hypotheses and reducing the impact of ASR errors on ER tasks.
Additionally, we designed a novel multimodal fusion (MF) module for learning frame-level MSRs and MIRs in a shared speech\textendash text modality space. 
We observed the following shortcomings in our previous work: 1) Despite introducing the AED and AEC modules, uncorrected ASR errors continue to affect the accuracy of SER. 2) Existing MF methods possibly fail to capture subtle differences in emotional features. 

To address these issues, we propose two learning strategies. First, we introduce adversarial learning to enhance the diversity and robustness of modality representations, thereby reducing the impact of ASR errors on SER tasks. Second, we introduce a contrastive learning strategy based on emotion labels to more accurately distinguish and capture feature differences between different emotion categories, thus improving the performance and robustness of the SER system.
In summary, our main contributions are as follows:

\begin{itemize}[leftmargin=*]
\vspace{-1mm}
\setlength{\topsep}{0pt}
\setlength{\itemsep}{0pt}
\setlength{\parsep}{0pt}
\setlength{\parskip}{0pt}

\item We introduce an adversarial modality discriminator to enhance the diversity of MSRs and improve the generalization of MIRs. This design alleviates modality mismatch and suppresses modality-specific noise introduced by ASR errors, which often cause misalignment between acoustic and textual features and degrade the robustness of multimodal fusion.

\item We propose a label-based contrastive learning strategy to enhance the discriminability of emotion representations across modalities. By encouraging emotion instances of the same category to cluster more tightly and separating those from different categories, this approach addresses the problem of emotion category confusion caused by overlapping feature distributions, especially in complex multimodal scenarios.

\item On the basis of our previous work, we present M$^4$SER, an SER method that leverages multimodal, multirepresentation, multitask, and multistrategy learning. The difference between our M$^4$SER and previous SER methods is illustrated in Fig. \ref{fig:pipeline}. M$^4$SER mitigates the impact of ASR errors, improves modality-invariant representations, and captures commonalities between modalities to bridge their heterogeneity.
\item Our proposed M$^4$SER outperforms state-of-the-arts on the IEMOCAP and MELD datasets. Extensive experiments demonstrate its superiority in SER tasks.
% \vspace{-1mm}
\end{itemize}

The remaining sections of this paper are organized as follows: In Section \ref{related_work}, we review related work. In Section \ref{sec:method}, we discuss the specific model and method of M$^4$SER. In Section \ref{sec:setup}, we outline the experimental setup used in this study, including datasets, evaluation metrics, and implementation details. In Section \ref{sec:results}, we verify the effectiveness of the proposed model. In Section \ref{sec:conclusion}, we present our conclusion and future work.
\vspace{-4mm}

% ==========Related Work==========
\section{Related Work}
\label{related_work}

\subsection{Multimodal SER}
SER aims to identify the speaker’s emotional state from spoken utterances. Early SER models primarily rely on audio signals, extracting prosodic and acoustic features such as Mel-frequency cepstral coefficients, filter banks, or other handcrafted descriptors~\cite{ghosh2022mmer}. With the advent of deep learning, models based on RNNs, CNNs and Transformer architectures~\cite{chen2022speechformer, fan2022isnet, shi2024study, mi2024two} have achieved significant improvements by modeling complex temporal and hierarchical patterns in speech. More recently, the rise of SSL has led to the development of speech-based pretrained models such as wav2vec~\cite{schneider2019wav2vec}, HuBERT~\cite{hsu2021hubert}, and WavLM~\cite{chen2022wavlm}, which provide rich contextual representations and deliver SOTA performance on various SER benchmarks~\cite{li2025seenet}.

In parallel, text-based SER has emerged as a complementary paradigm, where textual transcripts—either manually curated or obtained via ASR—serve as input for emotion prediction. These approaches benefit from high-level semantic understanding and often leverage contextual modeling techniques such as RNNs or graph neural networks (GNNs)~\cite{bharti2022text}. However, textual information alone lacks prosodic and paralinguistic cues, limiting its effectiveness in capturing nuanced emotional states.

To address these limitations, multimodal SER often integrates both speech and text modalities, capturing both acoustic and semantic cues. A common setting involves using speech signals alongside textual transcripts (typically from manual annotations or ASR outputs)~\cite{fan2023mgat, sun2024fine, fan2025coordination, he2025gia}.
Fan et al. \cite{fan2023mgat} proposed multi-granularity attention-based transformers (MGAT) to address emotional asynchrony and modality misalignment. 
Sun et al. \cite{sun2024fine} introduced a method integrating shared and private encoders, projecting each modality into a separate subspace to capture modality consistency and diversity while ensuring label consistency in the output. 
These methods typically assume access to high-quality manual transcriptions for the text modality.

Recently, growing efforts have focused on reducing the dependency on annotated text by directly leveraging ASR hypotheses as input \cite{santoso2021speech, lin2023robust, li2024speech, li2024crossmodal, li2025revise}. Santoso et al.~\cite{santoso2021speech} introduced confidence-aware self-attention mechanisms that downweight unreliable ASR tokens to mitigate error propagation. Lin and Wang~\cite{lin2023robust} proposed a robust multimodal SER framework that adaptively fuses attention-weighted acoustic representations with ASR-derived text embeddings, compensating for recognition noise in the transcript.

\vspace{-4mm}

\subsection{Multirepresentation: Modality-specific and Modality-invariant Representations}
 
Learning modality-specific representations (MSRs) and modality-invariant representations (MIRs) has proven effective for multimodal SER, as it enables models to capture both shared semantics across modalities and complementary modality-unique cues.
Hazarika et al.~\cite{hazarika2020misa} proposed MISA, which disentangles each modality into invariant and specific subspaces through factorization, facilitating effective fusion for emotion prediction.
Liu et al.~\cite{liu2024contrastive} proposed a modality-robust MER framework that introduces a contrastive learning module to extract MIRs from full-modality inputs, and an imagination module that reconstructs missing modalities based on the MIRs, enhancing robustness under incomplete input conditions.
To explicitly reduce distribution gaps and mitigate feature redundancy, Yang et al.~\cite{yang2022disentangled} proposed FDMER, which extracts both MSRs and MIRs, combined with consistency–disparity constraints and a modality discriminator to encourage disentangled learning.
Liu et al.~\cite{liu2024noise} presented NORM‑TR, introducing a noise‑resistant generic feature (NRGF) extractor trained with noise‑aware adversarial objectives, followed by a multimodal transformer that fuses NRGFs with raw modality features for robust representation learning.

In contrast to factorization-based approaches which explicitly disentangle modality inputs into shared and private spaces, our M\textsuperscript{4}SER framework adopts a structured encoding–generation strategy to implicitly model MSRs and MIRs. Specifically, we utilize a stack of cross-modal encoder blocks to extract token-aware and speech-aware features (MSRs), and then apply a dedicated generator composed of hybrid-modal attention and masking mechanisms to derive the MIRs. In addition, unlike prior works that incorporate adversarial learning at a global or representation-level, our M\textsuperscript{4}SER employs a frame-level discriminator that operates on each temporal representation. A frame-level discriminator attempts to distinguish the modality origin of each temporal representation, while the MIR generator is optimized to deceive the discriminator by producing modality-agnostic outputs that are hard to classify as either text or speech. This fine-grained, per-frame adversarial constraint leads to more robust and aligned cross-modal representations. Moreover, M\textsuperscript{4}SER integrates this disentanglement framework into a multitask learning setup with ASR-aware auxiliary supervision, enabling more effective adaptation to noisy real-world speech–text scenarios.

\vspace{-4mm}

\subsection{Multitask: ASR Error Detection and Correction}
ASR error correction (AEC) techniques remain effective in optimizing the ASR hypotheses. AEC has been jointly trained with ER tasks \cite{li2024crossmodal, lin2023robust, 10446548} in multitask settings. Li et al. \cite{li2024crossmodal} proposed an AEC method that improves transcription quality on low-resource out-of-domain data through cross-modal training and the incorporation of discrete speech units, and they validated its effectiveness in downstream ER tasks. Lin and Wang \cite{lin2023robust} introduced a robust ER method that leverages complementary semantic information from audio, adapts to ASR errors through an auxiliary task for ASR error detection, and fuses text and acoustic representations for ER. 
On the basis of \cite{DBLP:journals/corr/abs-2208-04641, he2025pmf, he2023enhancing}, we propose a partially autoregressive AEC method that uses a label predictor to restrict decoding to only desirable parts of the input sequence embeddings, thereby reducing inference latency.

\vspace{-4mm}

\subsection{Multistrategy Learning}
\subsubsection{Adversarial Network}
The concept of adversarial network was first introduced using GAN \cite{goodfellow2014generative}, which rapidly attracted extensive research interest owing to their strong capability to generate high-quality novel samples on the basis of existing data.
As research progressed, the applications of GAN expanded to multimodal tasks. 
In a multimodal SER task, Ren et al. \cite{ren2023maln} combined speaker characteristics with single-modal features using an adversarial module to capture both the commonality and diversity of single-modal features, and they finally fused different modalities to generate refined utterance representations for emotion classification.

\subsubsection{Supervised Contrastive Learning}
In recent years, to fully utilize supervised information, some researchers have proposed supervised contrastive learning (SCL) \cite{khosla2020supervised}, which leverages label information to construct positive pairs, making the distance between samples of the same class closer than that between samples of different classes. 
Zhang et al. \cite{zhang2022label} introduced label embeddings to better understand the language. Li et al. \cite{li2022targeted} unified the target setting on a hypersphere and forced the data representations to be close to these targets. Zhu et al. \cite{zhu2022balanced} regarded classifier weights as prototypes in the representation space and incorporated them into the contrastive loss. 
% ==========Related Work==========

% =======overall_architecture=======================
\begin{figure*}[htbp]
  \centering
   \includegraphics[width=2\columnwidth]{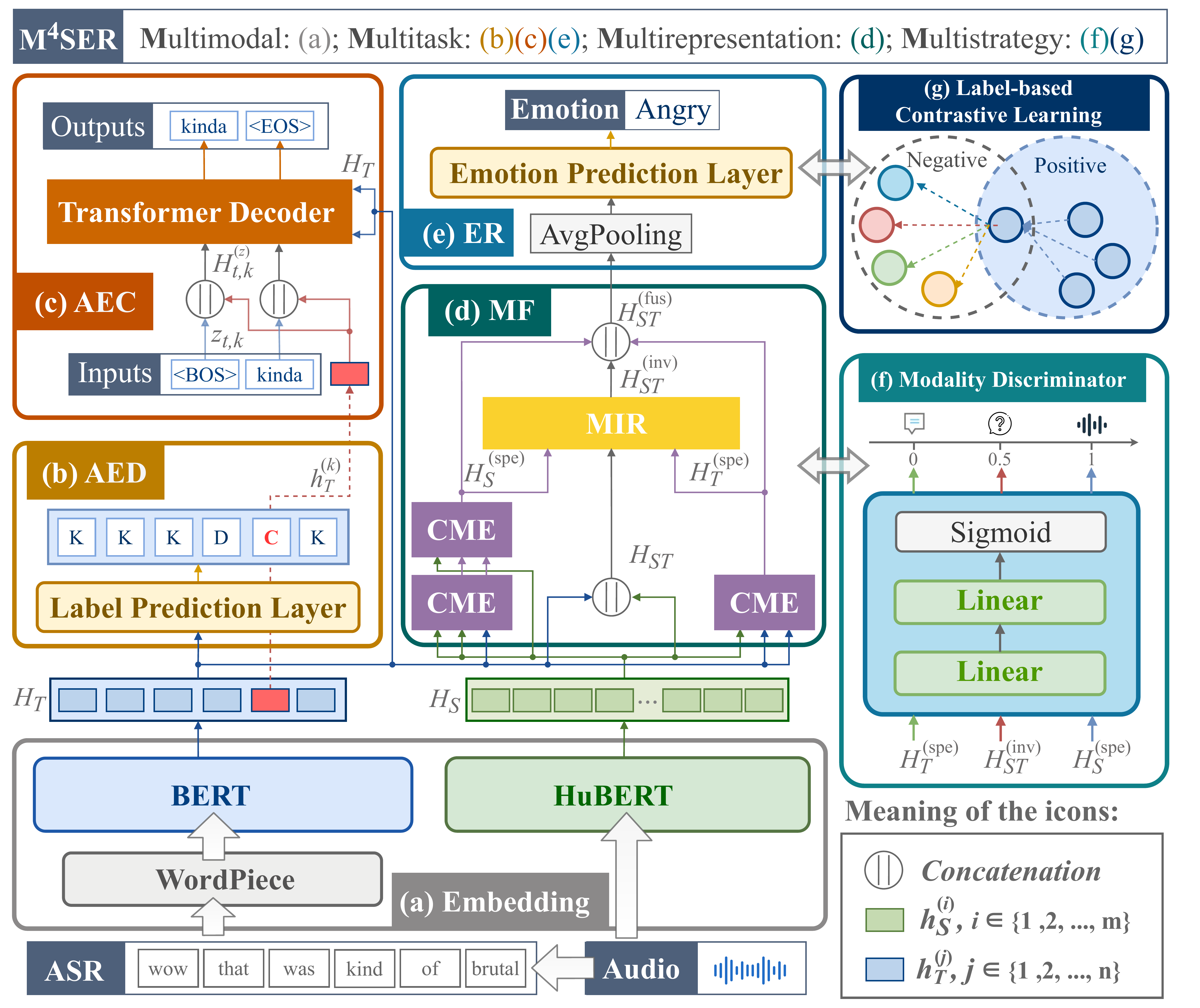}
  \vspace{-2mm}
  \caption{Overall architecture of the proposed M\textsuperscript{4}SER model. Specific illustrations of CME and MIR blocks in the (d) MF module are shown in Figs. \ref{fig:CME} and \ref{fig:MIR}, respectively.}
  \vspace{-4mm}
  \label{fig:model}
\end{figure*}
% =======overall_architecture=======================
\vspace{-4mm}
\section{Methodology}
\label{sec:method}

\subsection{Problem Formulation}
\label{section2.1}

Our proposed M\textsuperscript{4}SER framework can be formalized as the function $f(S, T) = (L, Z)$. Here, the speech modality $S = (s_1, s_2, \cdots , s_m) \in \mathbb{R}^{m}$ consists of \textit{m} frames extracted from an utterance, whereas the text modality $T = (t_1, t_2, \cdots , t_n) \in \mathbb{R}^{n}$ comprises $n$ tokens from the original ASR hypotheses of the utterance. All tokens are mapped to a predefined WordPiece vocabulary \cite{wu2016google}. Additionally, within our multitask learning framework, the primary task is ER, yielding an emotion classification output $L \in \{l_1, l_2, \cdots , l_e\}$, where $e$ represents the number of emotional categories. Concurrently, the auxiliary tasks include ASR error detection (AED) and ASR error correction (AEC). The output of these auxiliary tasks is $Z \in \{Z_1, Z_2,\cdots, Z_k \}$, representing all the ground truth (GT) sequences where the ASR transcriptions differ from the human-annotated transcriptions.
$Z_k = (z_{1,k}, z_{2,k}, \cdots , z_{c,k}) \in \mathbb{R}^{c}$, denoting that the $k^{th}$ error position includes $c$ tokens.

The overall architecture of the M\textsuperscript{4}SER model is
composed of five modules, namely, the embedding module, AED module, AEC module, MF module, and ER module, as shown in Fig. \ref{fig:model}.  In the following section, we provide a detailed explanation of each module.

\vspace{-5mm}
\subsection{Embedding Module}
\label{section2.2}
Our embedding module consists of acoustic and token embeddings, as illustrated in Fig. \ref{fig:model}(a).

\textbf{Contextual Speech Representations.} To obtain comprehensive contextual representations of acoustic features, we utilize a pretrained SSL model, HuBERT \cite{hsu2021hubert}, as our acoustic encoder. HuBERT integrates CNN layers with a transformer encoder to effectively capture both speech features and contextual information.
We denote the acoustic hidden representations of the speech modality inputs $S$ generated by HuBERT as $H_S = (h_S^{(1)}, h_S^{(2)}, \cdots , h_S^{(m)}) \in \mathbb{R}^{m \times d}$, where $d$ is the dimension of hidden representations.

% \noindent % 首行不缩进
\textbf{Contextual Token Representations.} We use the pretrained language model BERT \cite{devlin2018bert} as our text encoder to obtain the token representations $H_T = (h_T^{(1)}, h_T^{(2)}, \cdots , h_T^{(n)}) \in \mathbb{R}^{n \times d}$ for the text modality inputs $T$.

\vspace{-2mm}
\begin{equation}
    H_T = {\rm BERT}({\rm TE}(T)+{\rm PE}(T)) \in \mathbb{R}^{n \times d},
\label{eq3}
\end{equation}
\vspace{-1mm}
where ${\rm TE}(\cdot)$ and ${\rm PE}(\cdot)$ denote the token and position embeddings, respectively. 

\vspace{-5mm}
\subsection{ASR Error Detection (AED) Module}
The first subtask we introduce is AED, which is designed to detect the positions of ASR errors.
Similarly to \cite{DBLP:journals/corr/abs-2208-04641}, we align $T$ and $C$ by determining the longest common subsequence between them. The aligned tokens are labeled \textit{KEEP} (\textbf{K}), whereas the remaining tokens are labeled \textit{DELETE} (\textbf{D}) or \textit{CHANGE} (\textbf{C}). A specific example is illustrated in Fig. \ref{fig:model}(b).
The label prediction layer is a straightforward fully connected (FC) layer with three classes.
\begin{equation}
\label{eq2}
  P(y_o|h_T^{(o)}) = {\rm SoftMax}({\rm FC}(h_T^{(o)})) \in \mathbb{R}^{3},
\end{equation}
where $y_o$, $o \in \{1, \cdots, n\}$ is the predicted labeling operation. 
The label prediction loss is then defined as the negative log-likehood.
\begin{equation}
\label{equ:loss_d}
\mathcal{L}_{\rm AED} = - \sum_{o}{\rm log}(P(y_o|h_T^{(o)})).
\end{equation}

\vspace{-5mm}
\subsection{ASR Error Correction (AEC) Module}
We introduce a second subtask called AEC, which aims to correct the errors determined by AED.
Unlike conventional autoregressive decoders that start decoding from scratch, our decoder operates in parallel to the tokens predicted as \textbf{C}.

More specifically, once the erroneous positions are identified, the decoder constructs a separate generation sequence for each token labeled as \textbf{C}. Each sequence begins with a special start token $<$\textit{BOS}$>$ and generates the corrected text in an autoregressive manner. 
At each decoding step, the decoder input for a given correction sequence consists of all previously generated tokens, each concatenated with the representation of the corresponding erroneous token.
All correction sequences are processed in parallel by a shared transformer decoder, which attends to the full ASR representation $H_{T}$ as memory.

For the $k^{th}$ \textbf{C} position, the generated sequence of length $c$ by the transformer decoder can be represented as 
% $Z_k = (z_{1,k}, z_{2,k}, \cdots , z_{c,k}) \in \mathbb{R}^{c}$
$Z_k = z_{1:c,k} = (z_{1,k}, z_{2,k}, \cdots , z_{c,k})$
, where $z_{1,k}$ is initialized by a special start token $<$\textit{BOS}$>$. 
We compute the decoder input embeddings for the first $t$ steps as follows:
\begin{equation}
  H_{1:t,k}^{(z)} = {\rm FC}(({\rm TE}(z_{1:t,k})+{\rm PE}(z_{1:t,k})) \parallel h_T^{(k)})  \in \mathbb{R}^{t \times d} ,
\end{equation}
where ``$\parallel$'' denotes a concatenation operation along the feature dimension.
Here, $h_T^{(k)}$ is the representation of the $k^{th}$ erroneous token, repeated at each decoding position $1{:}t$.
Then, a generic transformer decoder \cite{vaswani2017attention} is applied to obtain the decoder layer output, where the query is $H_{1:t,k}^{(z)}$ and both the key and value are $H_T$:
\vspace{-4mm}

\begin{equation}
O_{t+1,k}^{{\rm(gen)}} = {\rm Transformer_{Decoder}}(H_{1:t,k}^{(z)},H_T,H_T)  \in \mathbb{R}^{d},
% % 
\vspace{-1mm}
\end{equation}
where $O_{t+1,k}^{\rm(gen)}$ is the decoder layer output. 
Finally, the generation output is calculated as

\begin{equation}
% \vspace{-2mm}
P_{t+1,k}^{\rm(gen)} = {\rm SoftMax}({\rm FC}(O_{t+1,k}^{\rm(gen)})) \in \mathbb{R}^{d_{vocab}},
\end{equation}
\vspace{-1mm}
where $d_{vocab}$ is the size of the vocabulary of BERT. Therefore, the next generated token is $z_{t+1,k} = {\rm argmax}(P_{t+1,k}^{\rm(gen)})$. Finally, the transformer-based generation loss is defined as
\begin{equation}
\begin{split}
\mathcal{L}_{\rm AEC} = - \sum_{k}\sum_{t}{\rm log}(P_{t,k}^{\rm(gen)}) .
\end{split}
\end{equation}

\vspace{-5mm}
\subsection{Multimodal Fusion (MF) Module}
On the basis of \cite{10446548}, our MF module is composed of three cross-modal encoder (CME) blocks and one MIR generator. The objective is to facilitate the learning of modality-specific and modality-invariant representations.
In this section, we provide an in-depth explanation of the operation of each CME block and the MIR generator.

\begin{figure}[t]
  \centering
  \includegraphics[width=0.5\columnwidth]{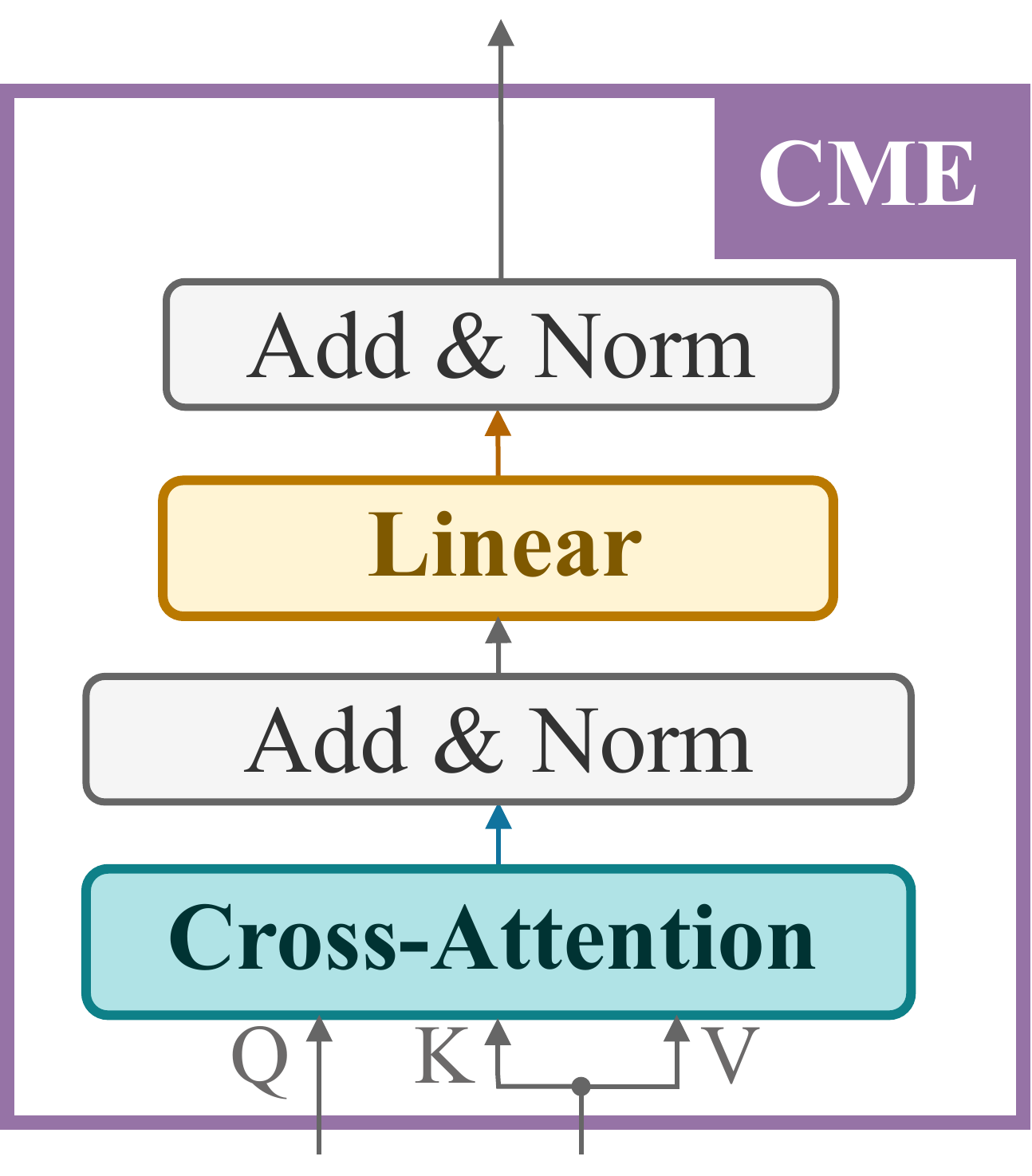}
  % \vspace{-6mm}
  \caption{Illustration of the CME block.}
  \vspace{-6mm}
  \label{fig:CME}
\end{figure}

\textbf{CME} is structured akin to a standard transformer layer, featuring an \textit{h}-head cross-attention module \cite{tsai2019multimodal}, residual connections, and FC layers, as shown in Fig. \ref{fig:CME}.
To acquire token-aware speech representations, we first feed $H_T$ as the query and $H_S$ as the key and value into a CME block:
\begin{equation}
\hat{H}_S^{(spe)} = {\rm FC}({\rm Cross \text{-}}{\rm Attn}(H_{T}, H_S, H_S)) \in \mathbb{R}^{n \times d},
% \vspace{-3mm}
\end{equation}
where ${\rm Cross \text{-}}{\rm Attn}$ is the cross-attention calculation.
Next, to address the issue where each representation generated in the previous block corresponds to token embeddings rather than acoustic embeddings, we input $\hat{H}_S^{(spe)}$ into another CME module. In this block, the original $H_S$ serves as the query and $\hat{H}_S^{(spe)}$ serves as the key and value. Ultimately, we obtain the final token-aware speech representation:
\begin{equation}
H_S^{{\rm(spe)}} = {\rm FC}({\rm Cross \text{-}}{\rm Attn}(H_{S}, \hat{H}_S^{(spe)}, \hat{H}_S^{(spe)})) \in \mathbb{R}^{m \times d}.
\end{equation}
Similarly, we feed $H_S$ as the query and $H_T$ as the key and value into a CME block to obtain speech-aware token representations:
\begin{equation}
H_T^{{\rm(spe)}} = {\rm FC}({\rm Cross \text{-}}{\rm Attn}(H_{S}, H_T, H_T)) \in \mathbb{R}^{m \times d}.
\end{equation}
% 

% =======CME=======================
% =======CME=======================

\begin{figure}[t]
  \centering
  \includegraphics[width=1\columnwidth]{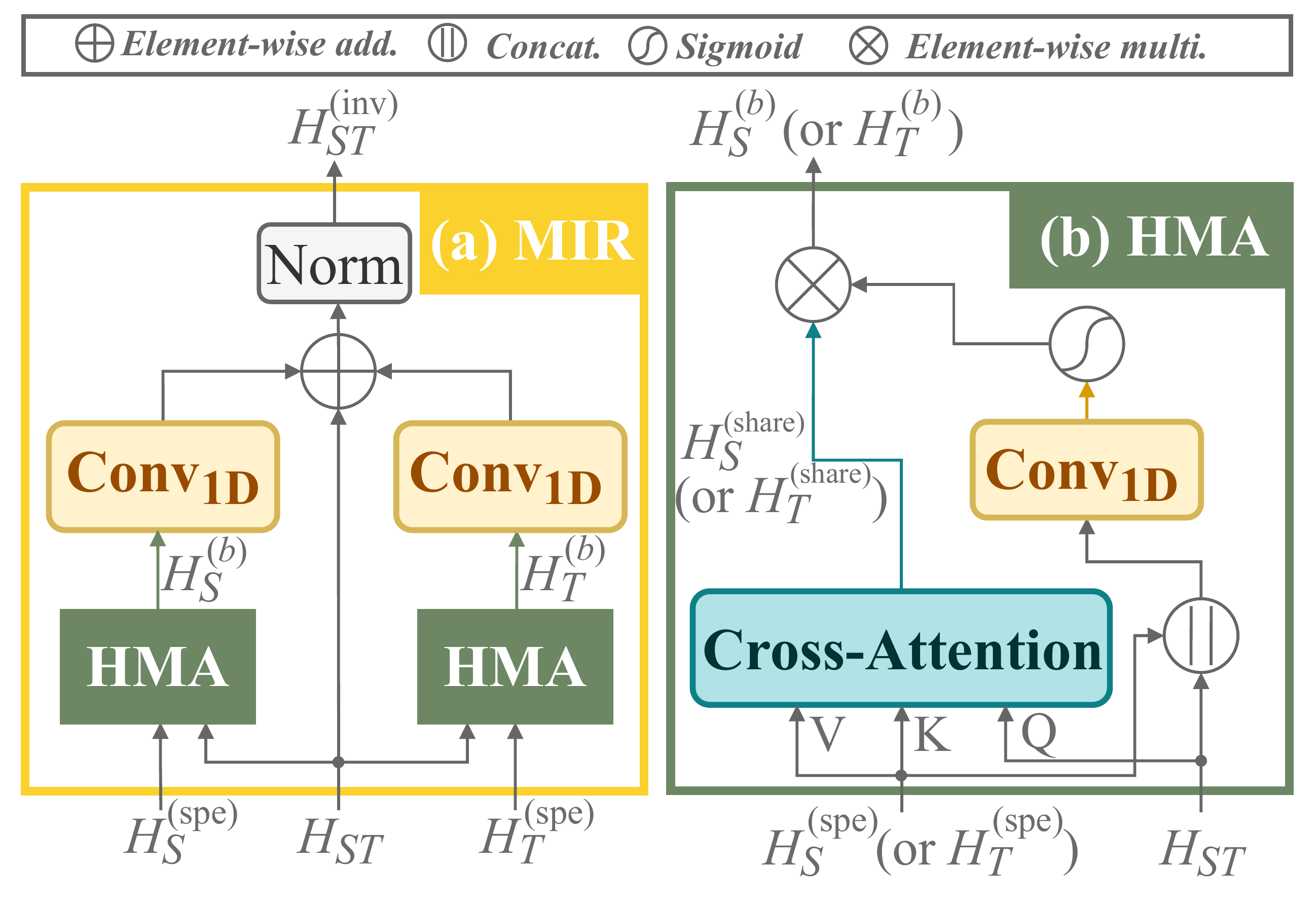}
  % \vspace{-6mm}
  \caption{Illustration of the MIR generator and HMA blocks.}
  \vspace{-6mm}
  \label{fig:MIR}
\end{figure}

The \textbf{MIR generator} utilizes a hybrid-modal attention (HMA) module to extract the shared information from each modality-specific representation that pertains to both modalities, as shown in Fig. \ref{fig:MIR}(a):
\begin{equation}
\label{equ:gen1}
H_i^{(b)} = {\rm HMA}( H_i^{\rm(spe)}, \text{FC}(H_{ST})) \in \mathbb{R}^{m \times d}, i \in \{S,T\},
\end{equation}
where $i$ denotes either the speech or text modality and $H_{ST}$ represents the concatenation of $H_{S}$ and $H_{T}$, which is then fed into a FC layer to map it back to the same dimension as $H_{S}$. This process integrates both speech and text information into a unified representation. The resulting features are subsequently summed with $H_{ST}$, culminating in the ultimate modality-invariant representation:
\begin{equation}
\label{equ:gen2}
H_{ST}^{\rm(inv)} = {\rm Norm}( H_{ST} + \sum_{i \in \{S,T\}} {\rm Conv_{1d}}(H_i^{(b)})) \in \mathbb{R}^{m \times d},
\end{equation}
where “{\rm Norm}” represents layer normalization \cite{ba2016layer} and “${\rm Conv_{1d}}$” denotes $1 {\rm \times} 1$ convolution followed by PReLU activation \cite{he2015delving}.
We uniformly express Eqs. (\ref{equ:gen1}) and (\ref{equ:gen2}) as
\begin{equation}
\label{equ:gen_all}
H_{ST}^{\rm(inv)} = G( H_S^{\rm(spe)}, H_T^{\rm(spe)}, H_{ST} ) \in \mathbb{R}^{m \times d},
\end{equation}
where $G$ is the MIR generator.

\textbf{HMA} initiates with a cross-attention layer aimed at extracting the shared information from each modality-specific representation pertinent to both modalities, as shown in Fig. \ref{fig:MIR}(b):
\begin{equation}
\begin{aligned}
H_i^{\rm(share)} = {\rm Cross \text{-}}{\rm Attn}&(H_{ST},  H_i^{\rm(spe)}, H_i^{\rm(spe)}) \\
&\in \mathbb{R}^{m \times d}, \quad i \in \{S,T\}.
\end{aligned}
\end{equation}

To enhance the feature's modality invariance, a parallel convolutional network is employed to learn a mask that filters out modality-specific information:
\begin{equation}
\begin{aligned}
H_i^{(b)} = H_i^{\rm(share)} \otimes \sigma &({{\rm Conv_{1d}}(H_i^{\rm(spe)} \parallel H_{ST})}) \\
&\in \mathbb{R}^{m \times d}, \quad i \in \{S,T\},
\end{aligned}
\end{equation}
where “$\sigma$” denotes Sigmoid activation and “$\otimes$” indicates element-wise multiplication.
Finally, the modality-specific and modality-invariant representations are concatenated together to obtain the final multimodal fusion representation $H_{ST}^{\rm(fus)}$:

\begin{equation}
H_{ST}^{\rm(fus)} = H_S^{\rm(spe)} \parallel H_T^{\rm(spe)} \parallel H_{ST}^{\rm(inv)}\in \mathbb{R}^{3m \times d}.
\end{equation}
% 

% =======MIR=======================
% =======MIR=======================

\vspace{-5mm}
\subsection{Emotion Recognition (ER) Module}
Emotion classification is performed by applying the temporal average pooling layer to the output feature $H_{ST}^{\rm (fus)}$ of the MF module, followed by an FC layer and a SoftMax activation function.

\vspace{-3mm}

\begin{equation}
\label{eq17}
  H_{ST}^{\rm(ap)} = {\rm AvgPooling}(H_{ST}^{\rm(fus)}) \in \mathbb{R}^{d},
\end{equation}
\begin{equation}
\label{eq18}
  P(y_{\rm emo}|H_{ST}^{\rm(fus)}) = {\rm SoftMax}({\rm FC}(H_{ST}^{\rm(ap)}) \in \mathbb{R}^{e},
\end{equation}
where $y_{\rm emo}$ is the predicted emotion classification and $e$ is the number of emotion categories. The corresponding loss function can be defined as
\begin{equation}
\mathcal{L}_{\rm ER} = - \sum{\rm log}(P(y_{\rm emo}|H_{ST}^{\rm(fus)})).
\end{equation}
% 
% \vspace{-3mm}

\subsection{Modality Discriminator}

We further develop the modality discriminator $D$ utilizing the modality-invariant representations produced by the MIR generator. This discriminator illustrated in Fig. \ref{fig:model}(f) is composed of two linear layers with a ReLU activation function in between, followed by a Sigmoid activation function. To enhance the MIR's capability to remain modality-agnostic, we employ adversarial learning techniques. 
Specifically, the discriminator $D$ outputs a scalar value between 0 and 1 for each frame, where values close to 0 indicate the text modality, values close to 1 indicate the speech modality, and values near 0.5 represent an ambiguous modality. 
Here, $D(H) \in \mathbb{R}^{m \times 1}, H \in \{ H^{(spe)}_{T},H^{(spe)}_{S},  H_{ST}^{\rm(inv)} \}$.
On one hand, we aim for the discriminator that correctly classifies frames in the modality-specific representations $H^{(spe)}_{T}$ and $H^{(spe)}_{S}$ as 0 and 1, respectively. On the other hand, to enhance the modality agnosticism of the modality-invariant representation $H_{ST}^{\rm(inv)}$ produced by the MIR generator, we aim for a discriminator that outputs values close to 0.5, indicating an ambiguous modality. With this design of the MIR generator and discriminator, we define a generative adversarial training objective, mathematically represented as
\begin{align}
\label{eq:gan}
\mathcal{L}_{\text{GAN}} &= \mathcal{L}_{D} + \mathcal{L}_{G} \notag \\
&= \mathbb{E} \left[ \log D(H_{S}^{\text{(spe)}}) + \log (1 - D(H_{T}^{\text{(spe)}})) \right] \notag \\
&\quad + \mathbb{E} \left[ -\log D(H_{ST}^{\text{(inv)}}) - \log (1 - D(H_{ST}^{\text{(inv)}})) \right],
\end{align}
where $H_{ST}^{\rm(inv)} = G( H_S^{\rm(spe)}, H_T^{\rm(spe)}, H_{ST} )$ is the output of the MIR generator as shown in Eq. (\ref{equ:gen_all}) and $\mathbb{E}$ denotes the expectation over all temporal frames in a batch.

\vspace{-4mm}
\subsection{Label-based Contrastive Learning (LCL)}
\label{sec:lcl}

% =======LCL=======================
\begin{figure}[t]
  \centering
    \includegraphics[scale=0.8]{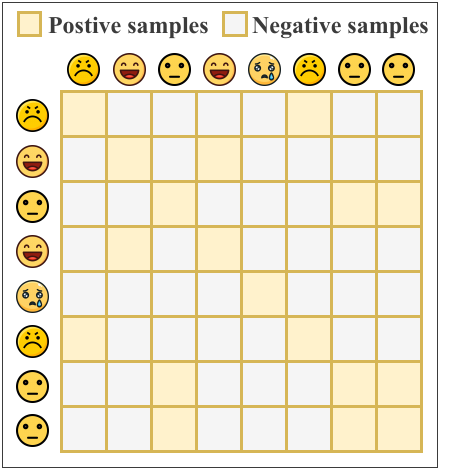}
  % \vspace{-6mm}
  \caption{Example of constructing positive and negative samples for label-based contrastive learning on the IEMOCAP dataset. In this example, the batch contains eight samples. Yellow squares represent positive samples, whereas white squares represent negative samples.}
  % \vspace{-6mm}
  \label{fig:LCL}
  \vspace{-5mm}
\end{figure}
% =======LCL=======================

To enhance the model's capability to learn emotion features from multimodal data, we employ a label-based contrastive learning task to complement the cross-entropy loss, as shown in Fig. \ref{fig:model}(g). This task aids the model in extracting emotion-related features when the MF module integrates speech and text data. As depicted in the LCL task in Fig. \ref{fig:LCL}, we categorize data in each batch into positive and negative samples on the basis of emotion labels. For instance, in a batch containing eight samples, we compute the set of positive samples for each sample, where those with the same label are considered positive samples (yellow squares), and those with different labels are considered negative samples (white squares). We then calculate the label-based contrastive loss (LCL) using Eq. \ref{eq:lcl}, which promotes instances of a specific label to be more similar to each other than instances of other labels.
\vspace{-4mm}

\begin{equation}
\begin{aligned}
\label{eq:lcl}
& \mathcal{L}_{\text{LCL}} = \sum_{i \in I} \mathcal{L}_{\text{LCL},i} \\
& = -\frac{1}{|P(i)|} \sum_{p \in P(i)} \log \frac{\exp(\langle (H_{ST}^{\rm(ap)})_i ,(H_{ST}^{\rm(ap)})_p \rangle / \tau)}{\sum_{a \in A(i)} \exp(\langle (H_{ST}^{\rm(ap)})_i , (H_{ST}^{\rm(ap)})_a \rangle / \tau)},
\end{aligned}
\end{equation}
where $\langle \cdot , \cdot \rangle$ denotes cosine similarity and \( \tau \) is the temperature parameter. For the multimodal representations, let $(H_{ST}^{\rm(ap)})_{i}$, $(H_{ST}^{\rm(ap)})_{p}$, and $(H_{ST}^{\rm(ap)})_{a}$ represent the $i^{th}$ sample, the $p^{th}$ positive sample, and the $a^{th}$ sample, respectively. We define \( I \) as the index set of samples, \( P(i) \) as the set of positive samples for the $i^{th}$ sample, and \( A(i) \) as the set of all samples.

\vspace{-4mm}
\subsection{Joint Training}

During the training stage, the learning process is optimized using three loss functions that correspond to ER, AED, and AEC (i.e., $\mathcal{L}_{\rm ER}$, $\mathcal{L}_{\rm AED}$, and $\mathcal{L}_{\rm AEC}$), and two training strategies (i.e., $\mathcal{L}_{\rm GAN}$ and $\mathcal{L}_{\rm LCL}$). The specific optimization process of M\textsuperscript{4}SER is detailed in Alg. \ref{alg:MIR}. 
These five loss functions are linearly combined as the overall training objective of M\textsuperscript{4}SER:
\begin{equation}
\mathcal{L} = \mathcal{L}_{\rm ER} + \alpha \cdot (\beta \cdot \mathcal{L}_{\rm AED} + \mathcal{L}_{\rm AEC}) + \gamma \cdot \mathcal{L}_{\rm GAN}+ \lambda \cdot \mathcal{L}_{\rm LCL},
\end{equation}
where $\alpha$ is the hyperparameter for adjusting the weight between the main task and the auxiliary tasks, and $\beta$ is the hyperparameter for adjusting the weight between the two auxiliary tasks, and $\gamma$, $\lambda$ are the hyperparameters for balancing different training strategies.

\begin{algorithm}
\caption{M\textsuperscript{4}SER Optimization}
\begin{algorithmic}[1]
\REQUIRE Training data $\mathcal{D}$ that contains multimodal inputs, namely, speech and corresponding ASR text pairs $(S, T)$, and outputs, namely, emotion labels, ASR operation labels, and the ground truth transcription $(y_{e}, y_{d}, y_{c})$. The M\textsuperscript{4}SER network $\theta$ that consists of Embedding $\theta_{Semb}$ and $\theta_{Temb}$, Encoders $\theta_{STenc}$, MIR generator $\theta_G$, modality discriminator $\theta_D$ and downstream emotion recognition module $\theta_{ER}$, ASR error detection module $\theta_{AED}$, and ASR error correction module $\theta_{AEC}$. Hyperparameter weights $\alpha$, $\beta$, $\gamma$, $\lambda$.
\STATE Randomly initialize the entire system $\theta$.
\IF{select self-supervised setting}
    \STATE Load the pretrained HuBERT model for $\theta_{Semb}$ and BERT model for $\theta_{Temb}$.
\ENDIF
\WHILE{not converged}
    \FOR{$(S, T) \in \mathcal{D}$}
        \STATE \textbf{Forward propagation:}
        \STATE $H_S = \theta_{Semb}(S), H_T = \theta_{Temb}(T)$ \hfill $\triangleright$ \textbf{\textcolor{blue!80!black}{Embedding}}
        \STATE $H^{{\rm (spe)}}_S, H^{{\rm (spe)}}_T = \theta_{STenc}(H_S, H_T)$ \hfill $\triangleright$ \textbf{\textcolor{blue!80!black}{CME}}
        \STATE $H_{ST} = H_S \parallel H_T$
        \STATE $H^{{\rm (inv)}}_{ST} = \theta_G(H^{{\rm (spe)}}_S, H^{{\rm (spe)}}_T, H_{ST})$ \hfill $\triangleright$ \textbf{\textcolor{blue!80!black}{Generator}}
        \STATE $\hat{y}_{e} = \theta_{ER}(H^{{\rm (spe)}}_S \parallel H^{{\rm (spe)}}_T \parallel H^{{\rm (inv)}}_{ST})$ \hfill $\triangleright$ \textbf{\textcolor{blue!80!black}{SER}}
        \STATE $\hat{y}_{d} = \theta_{AED}(H_T)$ \hfill $\triangleright$ \textbf{\textcolor{blue!80!black}{ASR Error Detection}}
        \STATE $\hat{y}_{c} = \theta_{AEC}(H_T)$ \hfill $\triangleright$ \textbf{\textcolor{blue!80!black}{ASR Error Correction}}
        \STATE \textbf{Training Objectives:}
        \STATE $\mathcal{L}_{\rm GAN} (\mathcal{L}_D \text{ and } \mathcal{L}_G) \text{ in Eq. \ref{eq:gan}}$ \hfill $\triangleright$ \textbf{\textcolor{blue!80!black}{Discriminator}}
        \STATE $\mathcal{L}_{\rm LCL} \text{ in Eq. \ref{eq:lcl} }$ \hfill $\triangleright$ \textbf{\textcolor{blue!80!black}{Contrastive learning}}
        \STATE $\mathcal{L}_{\rm ER} = \text{CrossEntropy}(\hat{y_e}, y_e)$
        \STATE $\mathcal{L}_{\rm AED} = \text{CrossEntropy}(\hat{y_d}, y_d)$
        \STATE $\mathcal{L}_{\rm AEC} = \text{CrossEntropy}(\hat{y_c}, y_c)$
        \STATE \textbf{Backpropagation:} \hfill $\triangleright$ \textbf{\textcolor{blue!80!black}{Adversarial training}}
        \STATE \textbf{Update Discriminator:} \hfill $\triangleright$ \textbf{\textcolor{blue!80!black}{unfreeze $\theta_D$}}
        \STATE $\theta_D \gets \arg\max \mathcal{L}_{\rm GAN}$
        \STATE \textbf{Update the Rest Network:} \hfill $\triangleright$ \textbf{\textcolor{blue!80!black}{freeze $\theta_D$}}
        \STATE $\theta \backslash \theta_D \gets \arg\min \mathcal{L}_{\rm ER} + \alpha \cdot (\beta \cdot \mathcal{L}_{\rm AED} + \mathcal{L}_{\rm AEC}) + \gamma \cdot \mathcal{L}_{G}+ \lambda \cdot \mathcal{L}_{\rm LCL}$
    \ENDFOR
\ENDWHILE
\end{algorithmic}
\label{alg:MIR}
\end{algorithm}

Following the GAN training strategy \cite{goodfellow2014generative}, we divide the backpropagation process into two steps. Firstly, we maximize $\mathcal{L}_{\rm GAN}$ to update the discriminator, during which the generator is detached from the optimization. According to Eq. (\ref{eq:gan}), on one hand, maximizing the first term $L_D$ of $\mathcal{L}_{\rm GAN}$ essentially trains the discriminator to correctly distinguish between text and audio features by making $H_S^{\rm(spe)}$ close to 1 and $H_T^{\rm(spe)}$ close to 0 \footnote{Function $\log(x) + \log(1 - y)$, where $x,y \in (0,1)$ reaches its maximum when $x$ approaches 1 and $y$ approaches 0.}. On the other hand, maximizing the second term $L_G$ (i.e., minimizing $-L_G$) will make $H^{\text{(inv)}}_{ST}$ approach 0 or 1 \footnote{Function $\log(x) + \log(1-x)$, where $x \in (0,1)$ reaches its maximum at $x = 0.5$ and its minimum near $x = 0$ and $x = 1$.}, indicating that the discriminator recognizes $H^{\text{(inv)}}_{ST}$ as modality-specific, which can be either text or audio, contrary to our desired modality invariance. Secondly, we freeze the discriminator parameters and update the remaining parameters by minimizing $L_G$, which makes the output of the discriminator $D$ for the modality-invariant features $H^{\text{(inv)}}_{ST}$ approach 0.5, thereby blurring these features between audio and text modalities to achieve modality agnosticism. Additionally, $\mathcal{L}_{\text{ER}}$ optimizes the downstream emotion recognition model, $\mathcal{L}_{\text{AED}}$ and $\mathcal{L}_{\text{AEC}}$ optimize the auxiliary tasks of ASR error detection and ASR error correction models, respectively, and $\mathcal{L}_{\text{LCL}}$ implements the LCL strategy. The entire system is trained in an end-to-end manner and optimized by adjusting weight parameters.

During the inference stage, the AED and AEC modules are excluded. The remaining modules accept speech and ASR transcripts as input and output emotion classification results.

\vspace{-1mm}
\section{Experimental Setup}
\label{sec:setup}
\vspace{-1mm}

In this section, we present the datasets, evaluation metrics, and implementation details used in our paper.

\vspace{-4mm}
\subsection{Dataset and Evaluation Metrics}
\label{ssec:Dataset}

% =======datasets=======================
\begin{figure}[t]
  \centering
  \includegraphics[width=1\columnwidth]{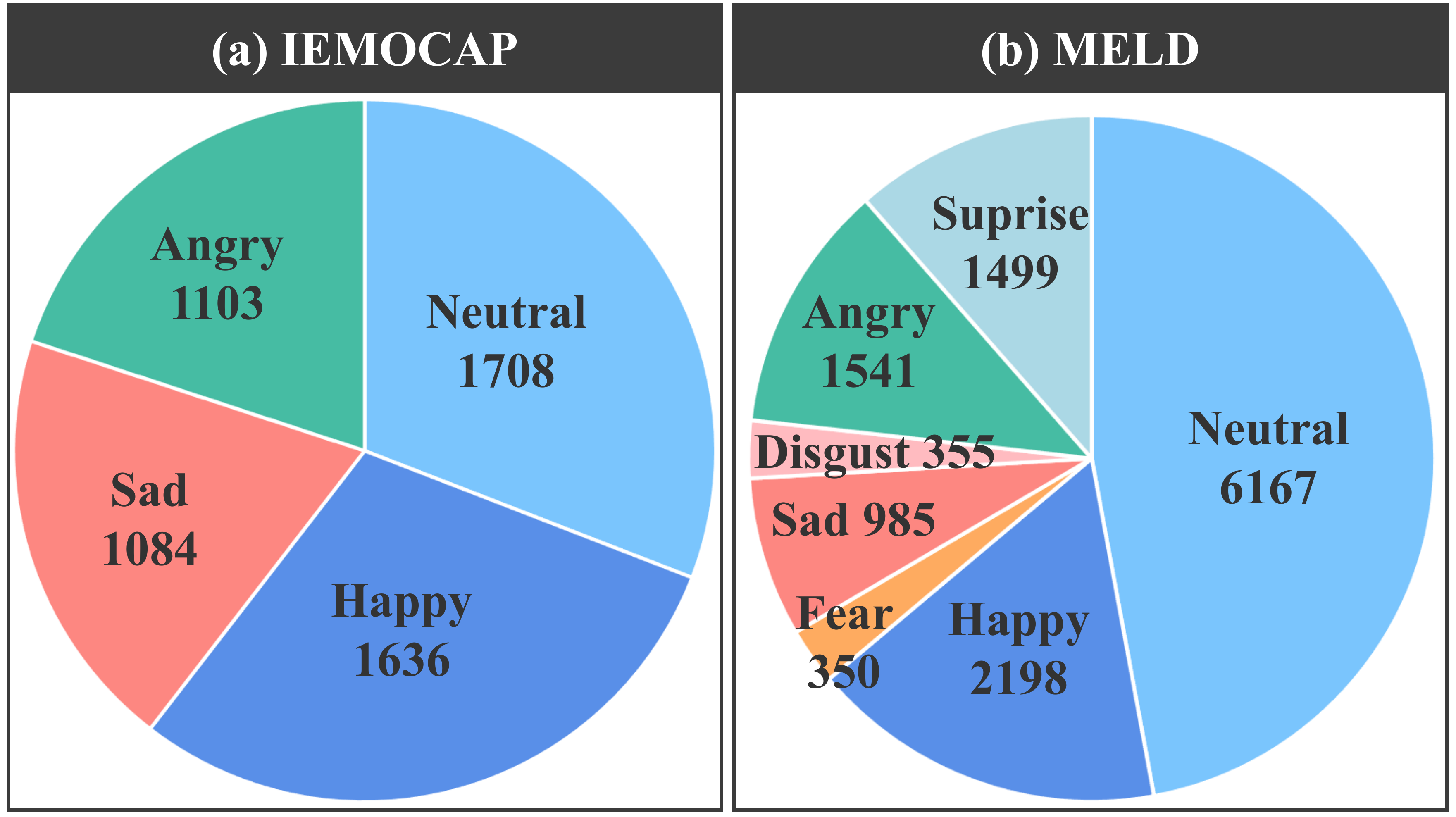}
  \vspace{-6mm}
  \caption{Amount of data for each type of emotion in IEMOCAP and MELD.}
  \vspace{-4mm}
  \label{fig:datasets}
\end{figure}
% =======datasets=======================

To investigate the effectiveness
of our proposed model, we carried out experiments on two public datasets: 
the Interactive Emotional Dyadic Motion Capture (IEMOCAP) \cite{busso2008iemocap} and the Multimodal Emotion Lines Dataset (MELD) \cite{poria2018meld}. The statistics of the two datasets are shown in Fig. \ref{fig:datasets}.

\textbf{IEMOCAP} comprised roughly 12 hours of speech from ten speakers participating in five scripted sessions. Following prior work \cite{dutta2022multimodal, lin2023robust} on IEMOCAP, we employed 5531 utterances that were categorized into four emotion categories: ``Neutral" (1,708), ``Angry" (1,103), ``Happy" (including ``Excited") (595 + 1,041 = 1,636), and ``Sad" (1,084). 
We performed five-fold leave-one-session-out (LOSO) cross-validation to evaluate the emotion classification performance using weighted accuracy (WA) and unweighted accuracy (UA).

\textbf{MELD} included a total of 13708 samples extracted from the TV series Friends, divided into 9,989 for training, 1,109 for validation, and 2,610 for testing. The dataset contained seven labels: ``Neutral" (6167), ``Happy" (2198), ``Fear" (350), ``Sad" (985), ``Disgust" (355), ``Angry" (1541), and ``Surprise" (1499). 
We tuned hyperparameters on the validation set and reported final results on the test set using the best checkpoint, evaluated by accuracy (ACC) and weighted F1-score (W-F1).

\vspace{-3mm}
\subsection{Implementation Details}
% \vspace{-3mm}

\label{ssec:Experiment Settings}

Our method was implemented using Python 3.10.0 and Pytorch 1.11.0. The model was trained and evaluated on a system equipped with an Intel(R) Xeon(R) Gold 6248 CPU @ 2.50GHz, 32GB RAM, and one NVIDIA Tesla V100 GPU. The detailed parameter settings are presented in Table \ref{tab:parameter_settings}.

% =======参数=====
\begin{table}[t]
\centering
\caption{Parameter Settings on the IEMOCAP and MELD Datasets.}
\label{tab:parameter_settings} % 添加标签
\begin{tabular}{lcc}
\toprule
\textbf{Parameter Settings} & \textbf{IEMOCAP} & \textbf{MELD} \\
\midrule
\midrule
Batch size & 16 & 16 \\
Epochs & 100 & 100 \\
Learning rate & $1e^{-5}$ & $1e^{-4}$ \\
Dropout & 0.5 & 0.5 \\
$\textit{d}_{S}$ & 768 & 768 \\
$\textit{d}_{T}$ & 768 & 768 \\
$\textit{d}_{vocab}$ & 30,522 & 30,522 \\
$\textit{d}$ & 768 & 768 \\
Attention layers & 12 & 12 \\
Attention heads & 12 & 12 \\
% Dimension-$c_{i}^{m}$ & 100 & 100 \\
% Dimension-$s_{i}^{m}$ & 100 & 100 \\
% Dimension-$z_{i}$ & 300 & 300 \\
$\alpha$ & 0.1 & 0.1 \\
$\beta$ & 3 & 3 \\
$\gamma$ & 0.01 & 0.01 \\
$\lambda$ & 0.1 & 0.1 \\
$\tau$ & 0.07 & 0.07 \\
\bottomrule
\end{tabular}
\vspace{-5mm}
\end{table}
% =======参数=====

The acoustic encoder was initialized using the \texttt{hubert-base-ls960}\footnote{\url{https://huggingface.co/facebook/hubert-base-ls960}} model, producing acoustic representations $d_S$ with a dimensionality of 768. 
The text encoder employed the \texttt{bert-base-uncased}\footnote{\url{https://huggingface.co/google-bert/bert-base-uncased}} model for initialization, yielding token representations $d_T$ with a dimensionality of 768. The vocabulary size for word tokenization $d_{vocab}$ was set to 30,522. We set the hidden size $d$ to 768, the number of attention layers to 12, and the number of attention heads to 12. Both the HuBERT and BERT models were fine-tuned during the training stage.
The transformer decoder in Fig. \ref{fig:model}(c) adopted a single-layer transformer with a hidden size of 768. Additionally, we obtained corresponding ASR hypotheses using the Whisper ASR model \cite{radford2022robust}. We used the \texttt{openai/whisper-medium.en}\footnote{\url{https://huggingface.co/openai/whisper-medium.en}} and \texttt{openai/whisper-large-v2}\footnote{\url{https://huggingface.co/openai/whisper-large-v2}} models, achieving word error rates (WERs) of 20.48\% and 37.87\% on the IEMOCAP and MELD datasets, respectively.

We used Adam \cite{kingma2014adam} as the optimizer with a batch size of 16. On the basis of empirical observations and prior work \cite{li2024enhancing, li2024cfn}, we kept the learning rate constant at $1e^{-5}$ for IEMOCAP and $1e^{-4}$ for MELD during training. For our multitask learning setup, we set $\beta$ to 3 and $\alpha$ to 0.1. For our multistrategy learning setup, we set $\gamma$ to 0.01 and $\lambda$ to 0.1. The temperature parameter $\tau$ was set to 0.07. To validate these choices, we additionally performed a one-at-a-time sensitivity analysis on the hyperparameters $\alpha$, $\beta$, $\gamma$, and $\lambda$, as presented in Section~\ref{sec:sense_nal} and illustrated in Fig.~\ref{fig:sense_anal}.

\vspace{-4mm}
\section{Experimental Results}
\label{sec:results}

\subsection{Comparisons of State-of-the-art (SOTA) Methods}

In the experiments, we compared various multimodal speech emotion recognition SOTA methods with our proposed method, M\textsuperscript{4}SER. 
The comparison was conducted by the following methods on the IEMOCAP dataset:

\begin{itemize}[leftmargin=*]
% \vspace{-2mm}
\setlength{\topsep}{0pt}
\setlength{\itemsep}{0pt}
\setlength{\parsep}{0pt}
\setlength{\parskip}{0pt}

\item \textbf{SAWC} \cite{dutta2022multimodal} adjusts importance weights using confidence measures, reducing ASR error impact by emphasizing the corresponding speech segments.
\item \textbf{RMSER-AEA} \cite{lin2023robust} uses complementary semantic information, adapting to ASR errors through an auxiliary task and fusing text and acoustic representations for SER.
\item \textbf{SAMS} \cite{hou2023semantic} leverages high-level emotion representations as supervisory signals to build a multi-spatial learning framework for each modality, enabling cross-modal semantic learning and fusion representation exploration.
\item \textbf{MGAT} \cite{fan2023mgat} tackles emotional asynchrony and modality misalignment problems by employing a multi-granularity attention mechanism.
\item \textbf{MMER} \cite{ghosh2022mmer} employs early fusion and cross-modal self-attention between text and acoustic modalities, and addresses three novel auxiliary tasks to enhance SER. For fairness, we selected the result that did not introduce augmented text for comparison.
\item \textbf{IMISA} \cite{wang2024inter} involves mapping a speech\textendash text pair to a shared subspace, extracting modality-specific features and employing contrastive learning to align features at a sample level.
\item \textbf{MAF-DCT} \cite{wu2024multi} introduces a dual approach utilizing SSL representation and spectral features for comprehensive speech feature extraction, along with a dual cross-modal transformer to handle interactions.
\item \textbf{FDRL} \cite{sun2024fine} integrates modality-shared and modality-private encoders, incorporating fine-grained alignment and disparity components to enhance modal consistency and diversity.
\item In \textbf{MF-AED-AEC} \cite{10446548}, as in our previous work, we consider two auxiliary tasks to improve semantic coherence in ASR text and introduce a MF method to learn shared representations.
\vspace{-3mm}
\end{itemize}

% -----------IEMCCAP performance--------

\begin{table}[ht]
\centering
\caption{Performance Comparison of SOTA Multimodal Models on IEMOCAP (\%). ``S" and ``T" represent Speech and Text modalities, respectively. ``ASR'' and ``GT" denote ASR and ground truth transcripts, respectively. \textbf{Bold} indicates the best result, whereas \underline{underline} signifies the second-best result.}
\label{tab:results_IEMOCAP}
\begin{tabular}{lcccc}
\toprule
\textbf{Method} & \textbf{Year} & \textbf{Modality} & \textbf{WA} & \textbf{UA} \\
\midrule
\midrule
SAWC \cite{dutta2022multimodal} & 2022 & S+T(ASR) & 76.6 & 76.8 \\
RMSER-AEA \cite{lin2023robust} & 2023 & S+T(ASR) & 76.4 & 76.9 \\
SAMS \cite{hou2023semantic} & 2023 & S+T(GT) & 76.6 & 78.1 \\
MGAT \cite{fan2023mgat} & 2023 & S+T(GT) & 78.5 & 79.3 \\
MMER \cite{ghosh2022mmer} & 2023 & S+T(GT) & 78.1 & \uline{\underline{79.8}} \\
IMISA \cite{wang2024inter} & 2024 & S+T(GT) & 77.4 & 77.9 \\
MAF-DCT \cite{wu2024multi} & 2024 & S+T(GT) & \uline{\underline{78.5}} & 79.3 \\
FDRL \cite{sun2024fine} & 2024 & S+T(GT) & 78.3 & 79.4 \\
MF-AED-AEC \cite{10446548} & 2024 & S+T(ASR) & 78.1 & 79.3 \\
\rowcolor[HTML]{EFEFEF} % 灰色
% \rowcolor[HTML]{D0E9F5} %浅蓝色
\textbf{M$^4$SER} & \textbf{2024} & \textbf{S+T(ASR)} & \textbf{79.2} & \textbf{80.1} \\
\bottomrule
\end{tabular}
\vspace{-3mm}
\end{table}
% -----------IEMCCAP performance--------

% -----------MELD performance--------
\begin{table}[ht]
\centering
\caption{Performance Comparison of SOTA Multimodal Models on MELD (\%). 
``V" represents Visual modality.
$^{*}$ indicates the result of training with GT texts and testing with ASR texts. We reimplement the method indicated by $^{\circ}$ on MELD and obtain the corresponding result.}
\label{tab:results_MELD}
\begin{tabular}{lcccc}
\toprule
\textbf{Method} & \textbf{Year} & \textbf{Modality} & \textbf{ACC} & \textbf{W-F1} \\
\midrule
\midrule
Full \cite{yang2022contextual} & 2022 & S+T(ASR) & - & 61.4 \\
HCAM$^{*}$ \cite{dutta2023hcam} & 2023 & S+T(ASR) & - & 50.2 \\
DIMMN \cite{wen2023dynamic} & 2023 & S+T(GT)+V & 60.6 & 58.0 \\
MER-HAN \cite{zhang2023multimodal} & 2023 & S+T(GT) & 62.9 & 60.2 \\
MLCCT \cite{gong2023multi} & 2023 & S+T(GT) & 63.2 & 62.4 \\
SAMS \cite{hou2023semantic} & 2023 & S+T(GT) & 65.4 & 62.6 \\
MF-AED-AEC$^{\circ}$ \cite{10446548} & 2024 & S+T(ASR) & \uline{\underline{65.5}} & \uline{\underline{64.1}} \\
\rowcolor[HTML]{EFEFEF} % 灰色
% \rowcolor[HTML]{D0E9F5} %浅蓝色
\textbf{M$^4$SER} & \textbf{2024} & \textbf{S+T(ASR)} & \textbf{66.5} & \textbf{66.0} \\
\bottomrule
\end{tabular}
\vspace{-2mm}
\end{table}

% -----------MELD performance--------

SAMS and MF-AED-AEC methods are also utilized as baselines for the MELD dataset. Additionally, the following methods were included for comparison:

\begin{itemize}[leftmargin=*]
% \vspace{-2mm}
\setlength{\topsep}{0pt}
\setlength{\itemsep}{0pt}
\setlength{\parsep}{0pt}
\setlength{\parskip}{0pt}

\item \textbf{Full} \cite{yang2022contextual} uses contextual cross-modal transformers and graph convolutional networks for enhanced emotion representations and modality fusion.

\item \textbf{HCAM} \cite{dutta2023hcam} combines wav2vec audio and BERT text inputs with co-attention and recurrent neural networks for multimodal emotion recognition.

\item \textbf{DIMMN} \cite{wen2023dynamic} fuses multimodal data using multiview attention layers, temporal convolutional networks, gated recurrent units, and memory networks to model dynamic dependencies and contextual information.

\item \textbf{MER-HAN} \cite{zhang2023multimodal} integrates local intramodal, cross-modal, and global intermodal attention mechanisms to effectively learn emotional features through a structured process.

\item \textbf{MLCCT} \cite{gong2023multi} involves feature extraction, interaction, and fusion using SSL embedding models, Bi-LSTM, cross-modal transformers,  and self-attention blocks.

\end{itemize}

% \subsection{Results and Analysis}
% \vspace{-1mm}
\label{ssec:Results and Analysis}

Tables \ref{tab:results_IEMOCAP} and \ref{tab:results_MELD} compare the performance of the M$^4$SER method with those of recent multimodal SER methods on the IEMOCAP and MELD datasets. Our proposed M$^4$SER achieves a WA of 79.2\% and a UA of 80.1\% on IEMOCAP, and an ACC of 66.5\% and a W-F1 of 66.0\% on MELD, outperforming other models and demonstrating the superiority of M$^4$SER. Specifically, on the IEMOCAP dataset, compared with the recent MAF-DCT, the M$^4$SER method shows a 0.7\% improvement in WA and a 0.8\% improvement in UA, reaching a new SOTA performance. These improvements stem from our M$^4$SER's capability to mitigate the impact of ASR errors, capture modality-specific and modality-invariant representations across different modalities, enhance commonality between modalities through adversarial learning, and excel in emotion feature learning with the LCL loss.

\begin{table}[t]
\centering
\caption{Ablation study on IEMOCAP and MELD datasets (\%). ``w/o" means ``without". ``Concat" denotes ``concatenation" operation.}
\label{tab:Ablation_study}
\begin{tabular}{l|cccc}
\toprule
\multirow{2}{*}{\textbf{Method}} & \multicolumn{2}{c}{\textbf{IEMOCAP}} & \multicolumn{2}{c}{\textbf{MELD}} \\
\cmidrule(r){2-3} \cmidrule(r){4-5}
& \textbf{WA} & \textbf{UA} & \textbf{ACC} & \textbf{W-F1} \\
\midrule
\midrule
\textbf{M$^4$SER (Full)} & \textbf{79.2} & \textbf{80.1} & \textbf{66.5} & \textbf{66.0} \\

\rowcolor[HTML]{EFEFEF} 
\midrule
\multicolumn{5}{c}{\textbf{A. Impact of Multimodalities}} \\
(1) only Speech Modality & 68.9 & 69.8 & 51.7 & 45.1 \\
(2) only Text Modality & 65.1 & 66.4 & 62.2 & 58.6 \\
(3) Speech \& Text + Concat  & 73.4 & 75.2 & 63.9 & 60.8 \\

\rowcolor[HTML]{EFEFEF} 
\midrule
\multicolumn{5}{c}
{\textbf{B. Impact of Multirepresentations}} \\
(1) w/o Modality-Specific (MSR) & 77.3 & 78.4 & 65.0 & 62.4 \\
(2) w/o Modality-Invariant (MIR) & 78.9 & 79.5 & 65.6 & 64.5 \\
(3) w/o MSR \& MIR& 76.5 & 77.9 & 64.8 & 61.7 \\

\rowcolor[HTML]{EFEFEF} 
\midrule
\multicolumn{5}{c}
{\textbf{C. Impact of Multitasks}} \\
 (1) w/o ASR Error Detection & 76.9 & 78.3 & 64.3 & 61.7 \\
 (2) w/o ASR Error Correction & 78.0 & 79.2 & 65.9 & 64.1 \\
  (3) w/o AED \& AEC & 77.3 & 78.6 & 65.1 & 62.6 \\

\rowcolor[HTML]{EFEFEF} 
\midrule
\multicolumn{5}{c}{\textbf{D. Impact of Multistrategies}} \\
(1) w/o GAN & 78.7 & 79.8 & 65.8 & 65.3 \\
(2) w/o LCL & 78.2 & 79.6 & 65.9 & 64.8 \\
(3) w/o GAN \& LCL & 78.1 & 79.3 & 65.5 & 64.1 \\

\bottomrule
\end{tabular}
\vspace{-4mm}
\end{table}

\vspace{-4mm}
\subsection{Ablation Study}

Our proposed M$^4$SER model is composed of four types of learning: multimodal learning, multirepresentation learning, multitask learning, and multistrategy learning. To determine the impact of different types of learning on the performance of the emotion recognition task, ablation experiments were conducted on the IEMOCAP and MELD datasets, as presented in Table \ref{tab:Ablation_study}.

\textbf{Impact of Multimodalities.} 
To assess the necessity of multimodality, we conduct single-modal comparison experiments involving text and speech modalities. These experiments operate under a baseline framework excluding the MF module, LCL loss, and GAN loss, precluding intermodal interaction.
As shown in Table \ref{tab:Ablation_study}(A), in the single-modal experiments, the performance of speech modality in the IEMOCAP dataset is better than that of text modality, whereas the reverse is observed in the MELD dataset. This divergence suggests that emotional information is more distinctly conveyed through speech in IEMOCAP, whereas MELD leans towards text for emotional expression.
Moreover, the multimodal baseline model, which simply concatenates speech and text features for recognition (see Table \ref{tab:Ablation_study} A(3)), significantly enhances performance compared with single-modal baselines, highlighting the essential role of multimodal information.

\begin{figure*}[ht]
\centering
\includegraphics[width=0.25\linewidth]{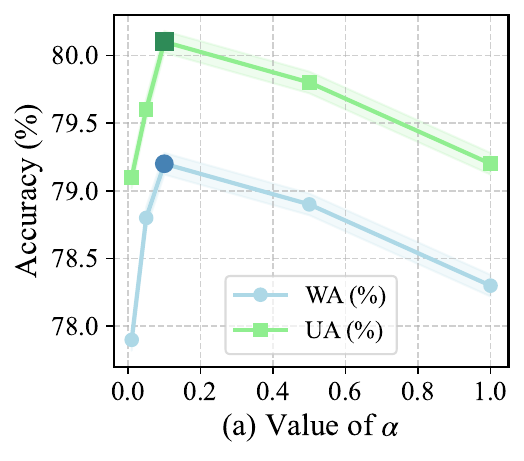}
\hspace{-3mm}
\includegraphics[width=0.25\linewidth]{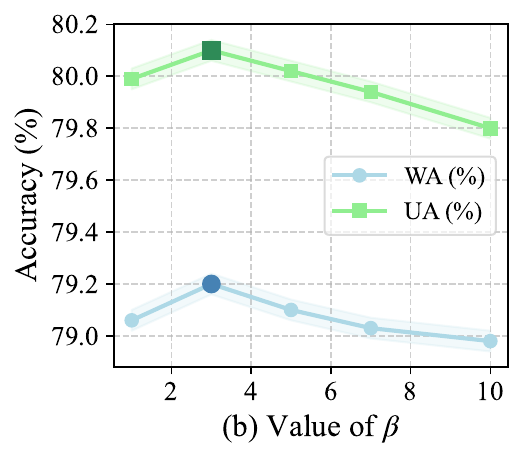}
\hspace{-3mm}
\includegraphics[width=0.25\linewidth]{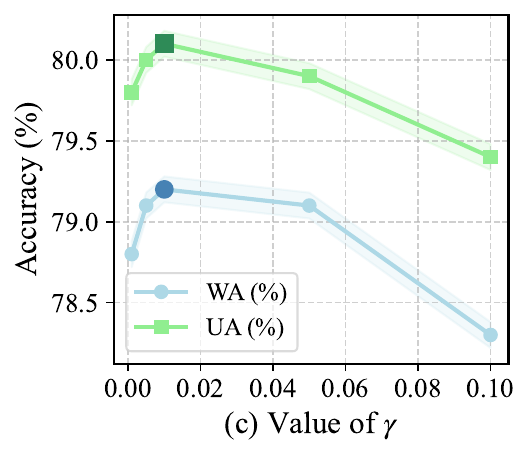}
\hspace{-3mm}
\includegraphics[width=0.25\linewidth]{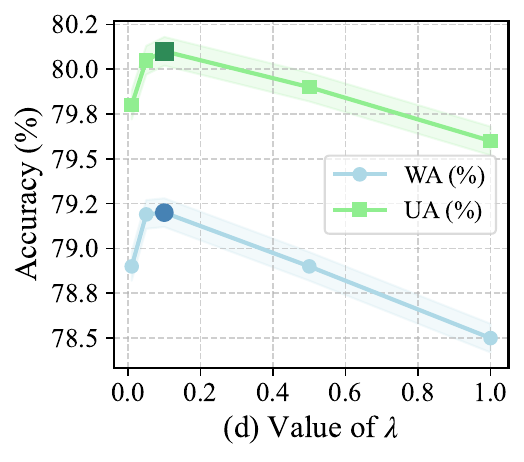}
\vspace{-3mm}
\caption{Sensitivity analysis of key hyperparameters on the IEMOCAP dataset.}
\label{fig:sense_anal}
\vspace{-3mm}
\end{figure*}

% =======loss=======================
\begin{figure*}[htbp]
  \centering
  \includegraphics[width=2\columnwidth]{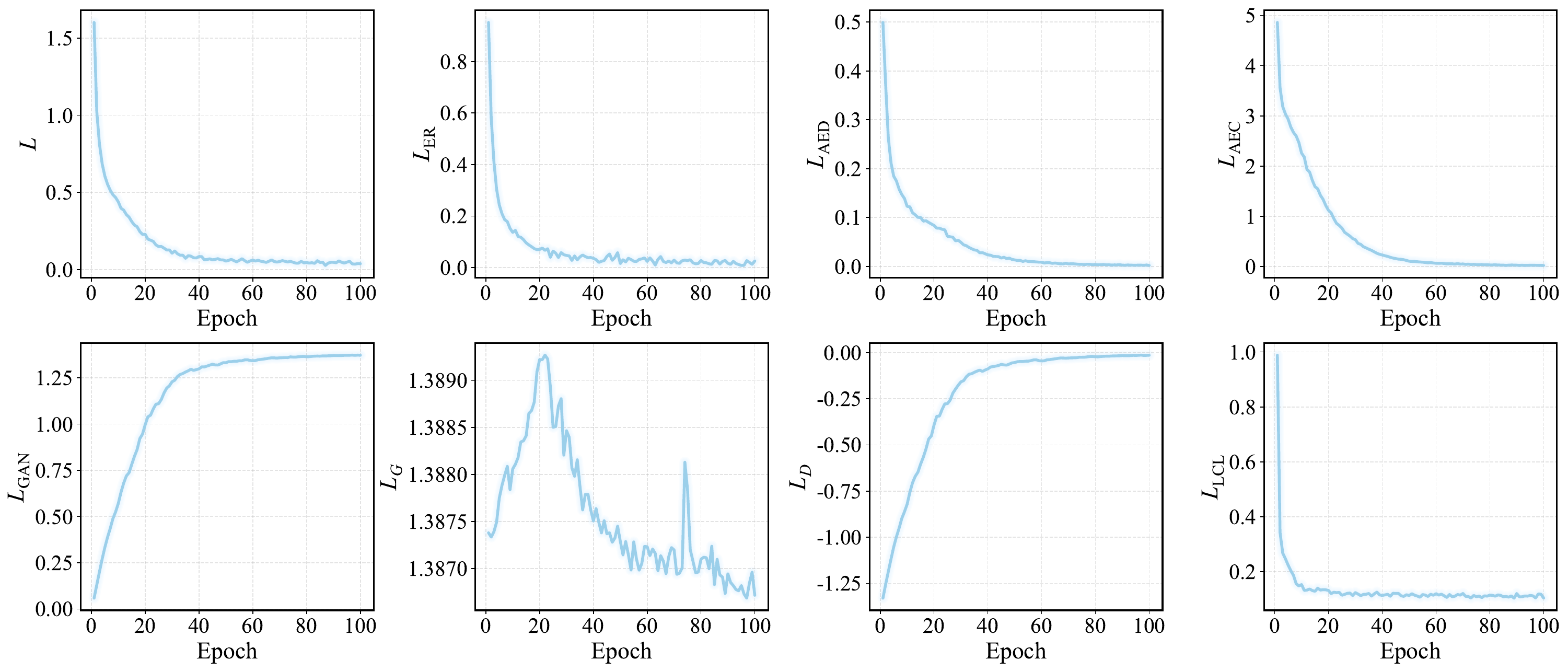}
  \vspace{-3mm}
  \caption{Evolution of optimization objectives during multitask multistrategy training.}
  \vspace{-4mm}
  \label{fig:loss_anal}
\end{figure*}
% =======loss=======================

\textbf{Impact of Multirepresentations.} We study the importance of modality-invariant and modality-specific representations by discarding each type. Removing modality-specific representations from multimodal fusion significantly reduces downstream emotion recognition performance on both datasets, confirming their effectiveness in capturing and utilizing unique information from each modality. Similarly, omitting refined modality-invariant representations from multimodal fusion also significantly decreases emotion recognition performance on both datasets, demonstrating their importance in bridging modality gaps. Clearly, removing both representations further decreases performance.

\textbf{Impact of Multitasks.} First, we discuss the impact of the AED module. Specifically, we remove the AED module and introduce only the AEC module as an auxiliary task. This requires the model to correct all errors in each utterance from scratch, rather than only the detected errors. Evidently, the absence of the AED module leads to a significant drop in emotion recognition performance, demonstrating the necessity of the AED module. Similarly, we observe that the AEC module also plays an important role. Moreover, we note that the WA results using only the AEC module (see Table \ref{tab:Ablation_study}(C)(1)) are even worse than the results with neither the AED nor AEC module (see Table \ref{tab:Ablation_study}(C)(3)) in both datasets. This phenomenon occurs because directly using the neural machine translation (NMT) model for AEC can even increase the WER \cite{leng2021fastcorrect}. Unlike NMT tasks, which typically require modifying almost all input tokens, AEC involves fewer but more challenging corrections. For instance, if the ASR model’s WER is 10\%, then only about 10\% of the input tokens require correction in the AEC model. However, these tokens are often difficult to recognize and correct because they have already been misrecognized by the ASR model. Therefore, we should consider the characteristics of ASR outputs and carefully design the AEC model, which is why we introduce both AED and AEC as auxiliary tasks.

% \vspace{-5mm}
\textbf{Impact of Multistrategies.} To validate the effectiveness of the adversarial training strategy outlined in Alg. \ref{alg:MIR}, we remove the adversarial training strategy from M$^4$SER. The results in Table \ref{tab:Ablation_study}(D)(1) show a decrease in performance on both datasets, demonstrating the critical role of this strategy in learning modality-invariant representations. The proposed modality discriminator effectively enhances the modality agnosticism of the refined representations from the generator. 
Additionally, omitting the LCL strategy described in Section \ref{sec:lcl} also results in similar performance degradation (see Table \ref{tab:Ablation_study}(D)(2)). We visualize the results before and after introducing this strategy to demonstrate its effectiveness, with specific analysis also detailed in the following t-SNE analysis section. Moreover, removing both of these strategies further diminishes emotion recognition performance, confirming that both learning strategies contribute significantly to performance improvement.

% =======tsne=======================
\begin{figure*}[htbp]
  \centering

  \includegraphics[width=2\columnwidth]{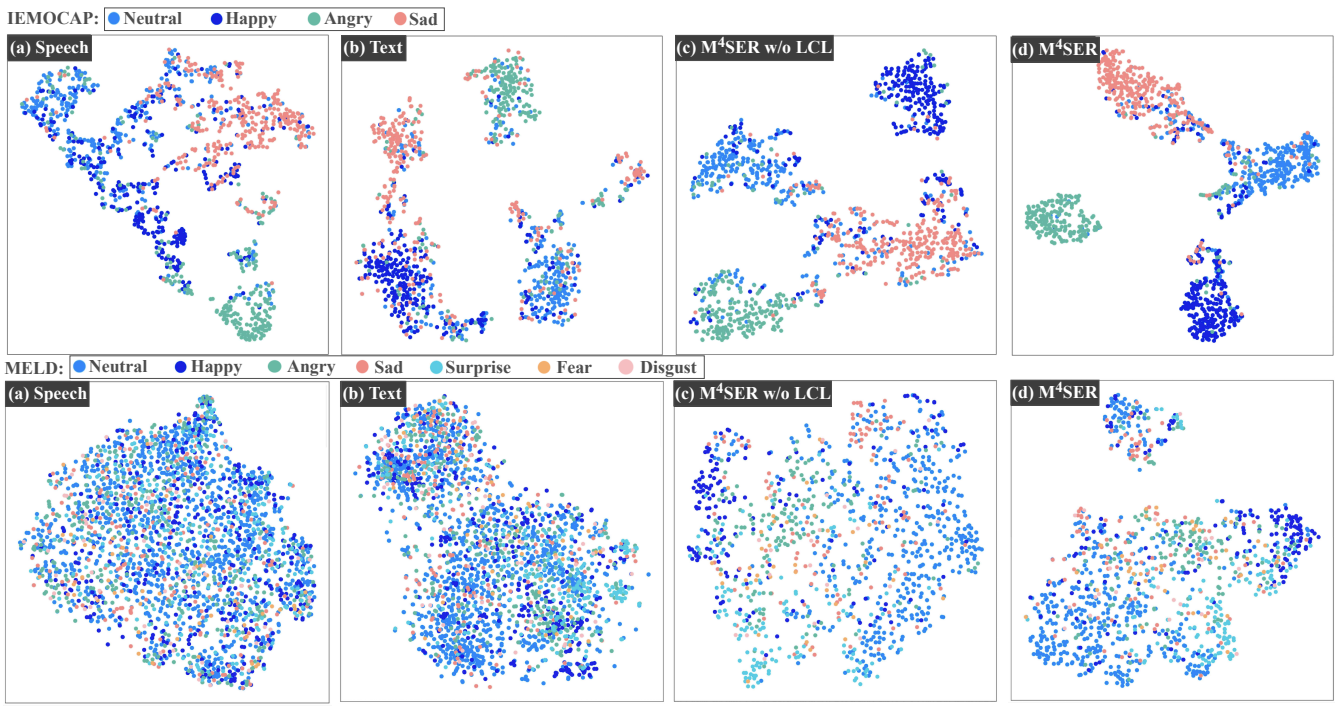}
  \vspace{-3mm}
  \caption{t-SNE visualization using IEMOCAP and MELD datasets. 
  We visualize all samples from IEMOCAP and MELD test sets. 
  }
  \vspace{-3mm}
  \label{fig:t-sne}
\end{figure*}
% =======tsne=======================

% =======gan=======================
\begin{figure}[t]
  \centering
    % \vspace{-6mm}
  \includegraphics[width=1\columnwidth]{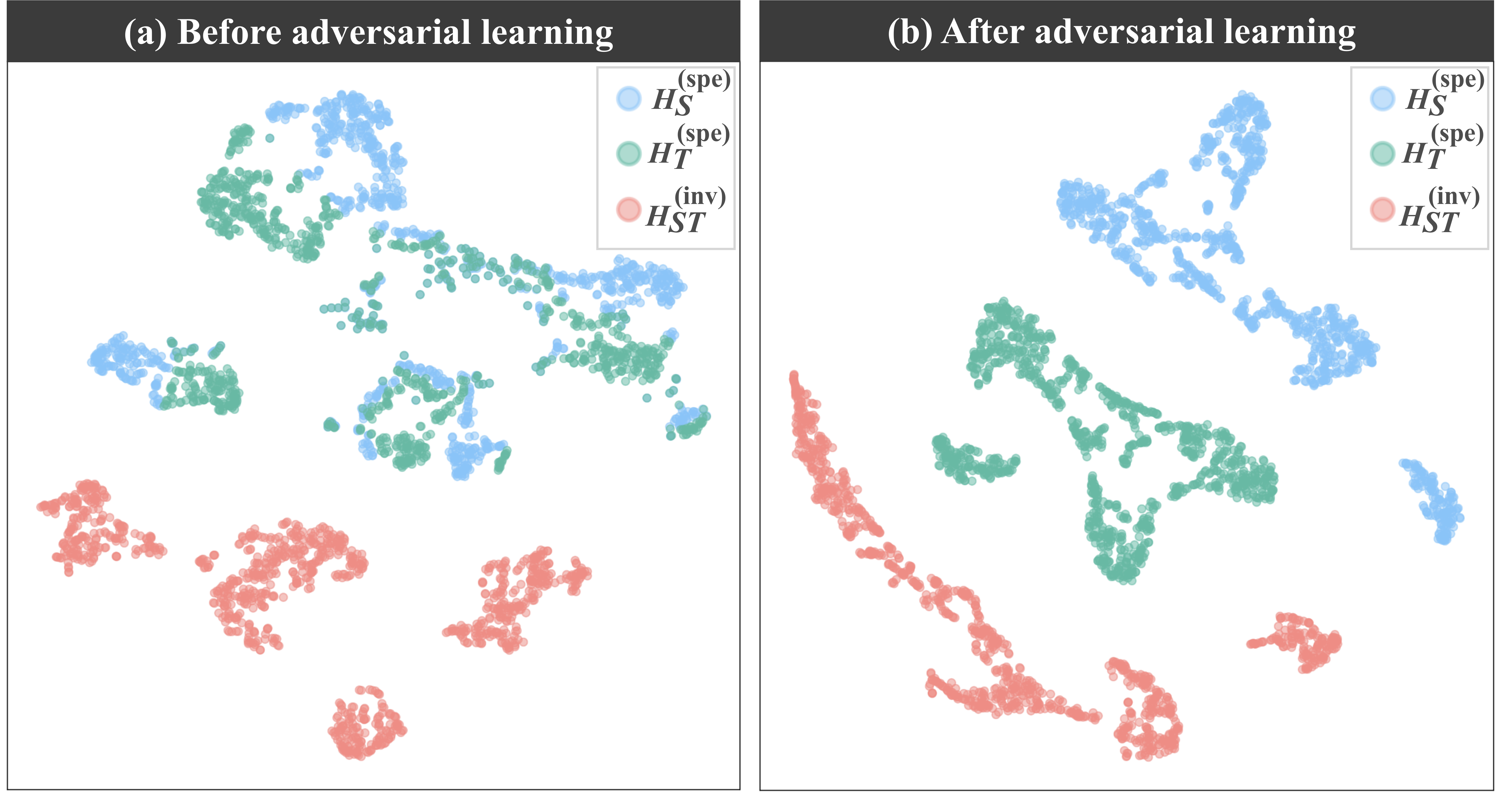}
  \vspace{-6mm}
  \caption{t-SNE visualizations of the distribution of the modality-specific and modality-invariant representations before and after adversarial learning on the IEMOCAP dataset.
  }
  \vspace{-6mm}
  \label{fig:gan}
\end{figure}

% =======gan=======================

% =======attention=======================
\begin{figure*}[htbp]
  \centering

  \includegraphics[width=2\columnwidth]{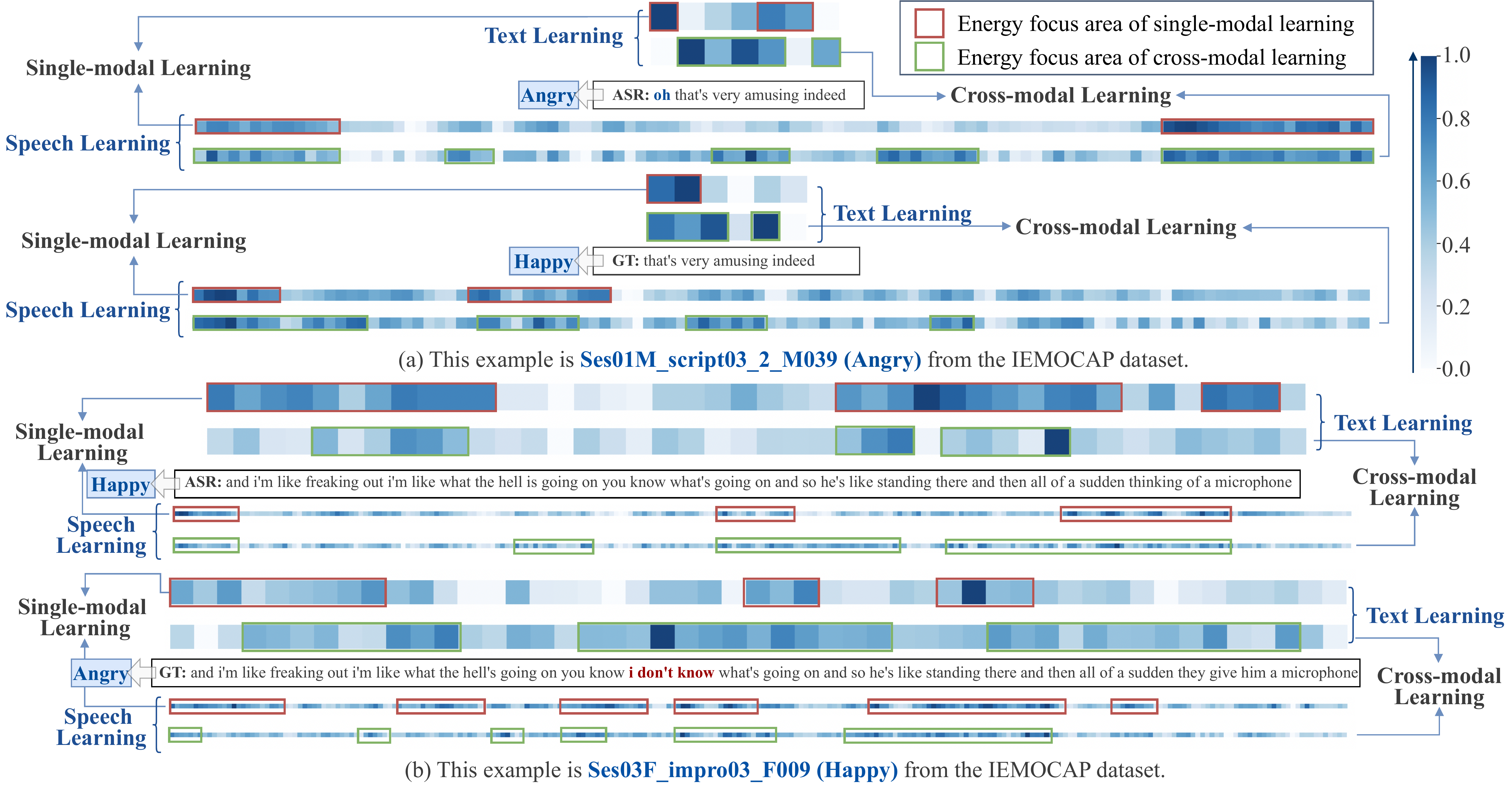}
  \vspace{-4mm}
  \caption{Representation weights of temporal-level features under different text input conditions in different types of modal learning. Brighter colors (tending towards blue) indicate higher values, suggesting the coverage of more key information.}

  \label{fig:attention_visualization}
  \vspace{-3mm}
\end{figure*}

% =======attention=======================

\begin{table*}[ht]
\centering
\caption{Trade-off between computational cost and performance on IEMOCAP. 
Parameter size (Params) is measured in millions (M). 
Training time is reported over 100 epochs in hours (h), and inference time is averaged per utterance in milliseconds (ms/U).}
\label{tab:cost-performance}
\begin{adjustbox}{max width=2\columnwidth}
\begin{tabular}{l|l|l|c|c|c|c|c}
\toprule
\textbf{Type} & \textbf{Model} & \textbf{Modality} & \textbf{Params (M)} & \textbf{Train Time (h)} & \textbf{Infer Time (ms/U)} & \textbf{WA (\%)} & \textbf{UA (\%)}  \\
\midrule
\midrule
\multirow{2}{*}{Single-modal} 
& HuBERT & S & 316.2 & 5.4 & 23.7 ± 0.1 & 68.9 & 69.8 \\
& BERT   & T(ASR)   & 109.5 & 0.5  & 6.3 ± 0.1 & 65.1 & 66.4 \\
\midrule
\multirow{4}{*}{Multi-modal}
& Baseline (HuBERT + BERT) & \multirow{4}{*}{S+T(ASR)} & 426.9 & 9.6 & 29.6 ± 0.1 & 73.4 & 75.2 \\
& \hspace{1em}+ MSR \& MIR & & 481.4 & 12.0 & 44.8 ± 0.1 & 76.6 & 77.4 \\
& \hspace{2em}+ GAN \& LCL &  & 481.6 & 20.8 & 44.5 ± 0.6 & 77.3 & 78.6 \\
& \cellcolor[HTML]{EFEFEF}\textbf{\hspace{3em}+ AED \& AEC (M$^4$SER) }   
& \cellcolor[HTML]{EFEFEF} 
& \cellcolor[HTML]{EFEFEF}\textbf{512.9} 
& \cellcolor[HTML]{EFEFEF}\textbf{21.5} 
& \cellcolor[HTML]{EFEFEF}\textbf{44.7 ± 0.2} 
& \cellcolor[HTML]{EFEFEF}\textbf{79.2} 
& \cellcolor[HTML]{EFEFEF}\textbf{80.1} \\
\bottomrule
\end{tabular}
\end{adjustbox}
\vspace{-3mm}
\end{table*}

\vspace{-5mm}
\subsection{Sensitivity Analysis}
\label{sec:sense_nal}

We conduct a one-at-a-time sensitivity analysis on four key hyperparameters: the auxiliary task weight $\alpha$, the AED/AEC balancing factor $\beta$, the GAN loss weight $\gamma$, and the LCL loss weight $\lambda$. As shown in Fig.~\ref{fig:sense_anal}, model performance is moderately sensitive to $\alpha$ and $\lambda$, with optimal results achieved at $\alpha=0.1$ and $\lambda=0.1$. Excessively small values reduce the effectiveness of auxiliary guidance, while overly large values interfere with the main task.

In contrast, the model exhibits relative robustness to variations in $\beta$, indicating stable synergy between the AED and AEC objectives. Conversely, performance is highly sensitive to $\gamma$; large values result in unstable training due to adversarial objectives, while a small weight (e.g., $\gamma=0.01$) ensures stable convergence and optimal results.

\vspace{-4mm}
\subsection{Convergence of Multitask and Multistrategy Losses}

Fig.~\ref{fig:loss_anal} shows that the multitask objectives 
($\mathcal{L}_{\rm ER}$, $\mathcal{L}_{\rm AED}$, $\mathcal{L}_{\rm AEC}$) 
rapidly decrease and stabilize, confirming that the auxiliary tasks provide 
useful supervision for ER. The multistrategy objectives also 
exhibit stable convergence: $\mathcal{L}_{\rm LCL}$ quickly decreases under label 
supervision, while for the adversarial component, $\mathcal{L}_{D}$ rises from 
negative values to near zero as the discriminator learns, $\mathcal{L}_{\rm GAN}$ 
steadily increases and plateaus, and $\mathcal{L}_{G}$ exhibits fluctuations, reflecting the adversarial interplay, 
and eventually converges as $D(H_{ST}^{\mathrm{(inv)}})\to 0.5$, which corresponds to about 1.386 in the $\mathcal{L}_{G}$ term of Eq.~(\ref{eq:gan}), 
indicating that an adversarial equilibrium is achieved. Overall, both multitask and multistrategy objectives 
are optimized in a stable manner, validating the effectiveness of the M$^4$SER.

\vspace{-3mm}
\subsection{Computational Complexity Analysis}

To assess the trade-off between model performance and computational cost, we report the number of parameters, training time, and inference time for each model on the IEMOCAP dataset in Table~\ref{tab:cost-performance}, with all experiments conducted on a single NVIDIA Tesla V100 GPU.

Among the single-modal models, HuBERT exhibits higher computational cost than BERT owing to the nature of speech encoders, but also delivers better recognition performance. In the multi-modal setting, each additional module introduces a moderate increase in parameter size and training cost, while consistently improving recognition accuracy.

Our full model M$^4$SER achieves the best performance with WA of 79.2\% and UA of 80.1\%. Although its training time increases to 21.5 hours owing to the inclusion of GAN and LCL strategies and AED and AEC subtasks, these components are used only during training. Consequently, the inference latency of M$^4$SER remains nearly identical to the baseline with only MSR and MIR. Considering the significant gains in performance, M$^4$SER demonstrates a favorable trade-off between computational cost and recognition accuracy, making it suitable for practical deployment.

\vspace{-3mm}
\subsection{Visualization Analysis}
\textbf{t-SNE Analysis.} To intuitively demonstrate the advantages of our proposed M$^4$SER model on IEMOCAP and MELD datasets, we utilize the t-distributed stochastic neighbor embedding (t-SNE) tool \cite{van2008visualizing} to visualize the learned emotion features using all samples from session 3 of the IEMOCAP and test set of the MELD. We compare these features across the following models: the speech model (Fig. \ref{fig:t-sne}(a)), the text model (Fig. \ref{fig:t-sne}(b)), the proposed M$^4$SER model without label-based contrastive learning (LCL) (Fig. \ref{fig:t-sne}(c)), and the proposed M$^4$SER model (Fig. \ref{fig:t-sne}(d)).

From the visualization results in Fig. \ref{fig:t-sne}, we observe that the emotion label distributions for the two single-modal baselines on the IEMOCAP and MELD datasets show a significant overlap among the emotion categories, indicating that they are often confused with each other. In contrast, the emotion label distributions for the two multimodal models are more distinguishable, demonstrating the effectiveness of multimodal models in capturing richer emotional features.

Furthermore, compared with the M$^4$SER model without the LCL strategy, our M$^4$SER model achieves better clustering for each emotion category, making them as distinct as possible. For instance, on the IEMOCAP dataset, the intra-class clustering of samples learned by the M$^4$SER model (Fig. \ref{fig:t-sne}(d)) is more pronounced than that learned by the M$^4$SER model without the LCL strategy (Fig. \ref{fig:t-sne}(c)), especially with clearer boundaries for sad and angry samples. Additionally, the distances between the four types of emotion in the emotion vector space increase. Although distinguishing MELD samples is more challenging than distinguishing IEMOCAP samples, the overall trend remains similar.
This indicates that the proposed M$^4$SER model effectively leverages both modality-specific and modality-invariant representations to capture high-level shared feature representations across speech and text modalities for emotion recognition.

To further explore the effectiveness of adversarial learning in M$^4$SER, we visualize the distribution of modality-invariant and modality-specific representations before and after adversarial learning using the t-SNE tool, as shown in Fig. \ref{fig:gan}. It can be observed that after adversarial learning, different modality-specific representations become more separated with increasing intraclass clustering. This indicates that with the introduction of adversarial learning, M$^4$SER enhances the diversity of modality-specific representations, generating superior features for the downstream emotion recognition task.

% -----------IEMCCAP performance--------

\begin{table}[t]
\centering
\caption{Performance comparison of our method on IEMOCAP using ASR and GT texts (\%). When using GT text, the AED and AEC modules are excluded.}
\label{tab:attn_analysis}
\begin{tabular}{lccc}
\toprule
\textbf{Method} & \textbf{Modality} & \textbf{WA} & \textbf{UA} \\
\midrule
\midrule
M$^4$SER & S+T(GT) & 78.4 & 79.9 \\

\rowcolor[HTML]{EFEFEF} % 灰色
% \rowcolor[HTML]{D0E9F5} %浅蓝色
\textbf{M$^4$SER} & \textbf{S+T(ASR)} & \textbf{79.2} & \textbf{80.1} \\
\bottomrule
\end{tabular}
\vspace{-3mm}
\end{table}
% -----------IEMCCAP performance--------

% -----------attention performance--------

\begin{table}[t]
    \centering
    \caption{Consistency Analysis between Cross-modal Speech and Text Representations Using ASR and GT Texts.}
    \label{tab:dtw}
    \begin{tabular}{lcccc}
        \toprule
        \multirow{2}{*}{\textbf{Metric}} & \multicolumn{2}{c}{\textbf{Example (a)}} & \multicolumn{2}{c}{\textbf{Example (b)}} \\
        \cmidrule(lr){2-3} \cmidrule(lr){4-5}
         & ASR  & GT  & ASR  & GT \\
        \midrule
        \midrule
        DTW $\downarrow$ & \textbf{0.52} & 1.89 & \textbf{1.46} & 1.57   \\
        \bottomrule
    \end{tabular}
\vspace{-5mm}
\end{table}

% -----------attention performance--------

\begin{figure*}[ht]
\centering
\includegraphics[width=0.25\linewidth]{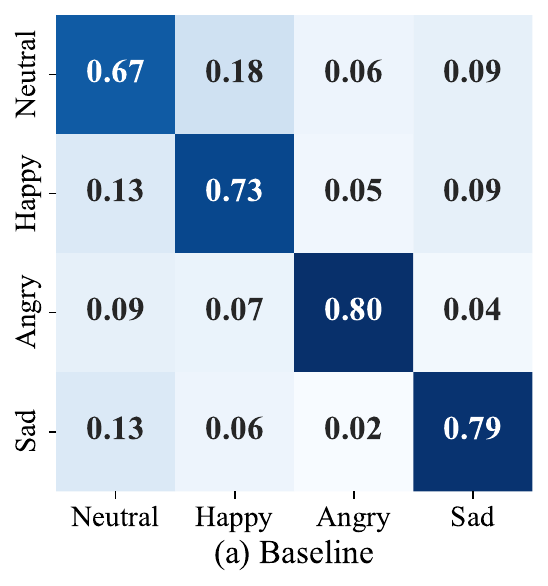}
\hspace{-3mm}
\includegraphics[width=0.25\linewidth]{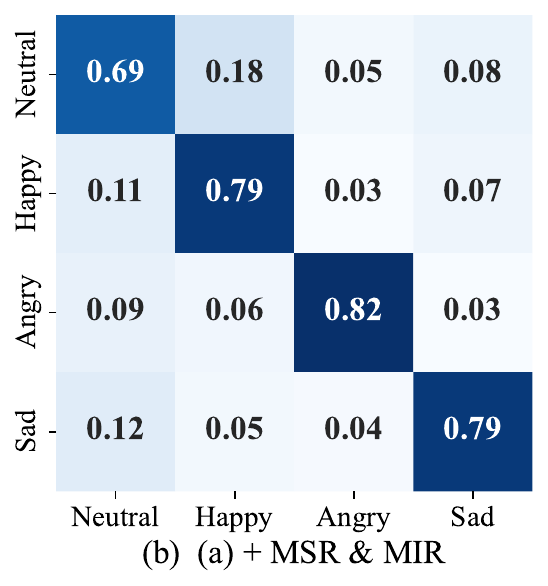}
\hspace{-3mm}
\includegraphics[width=0.25\linewidth]{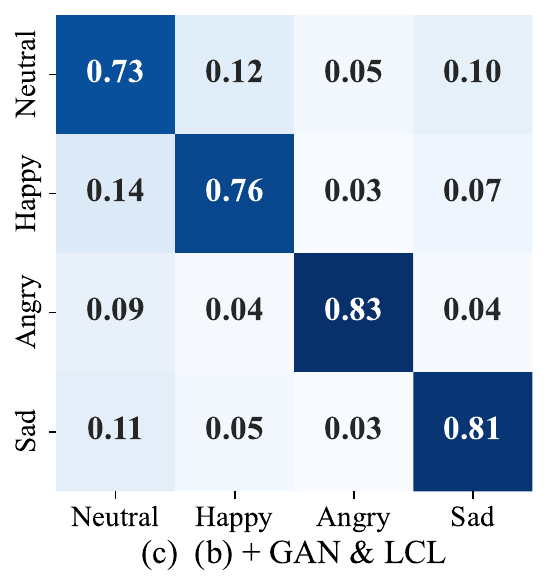}
\hspace{-3mm}
\includegraphics[width=0.25\linewidth]{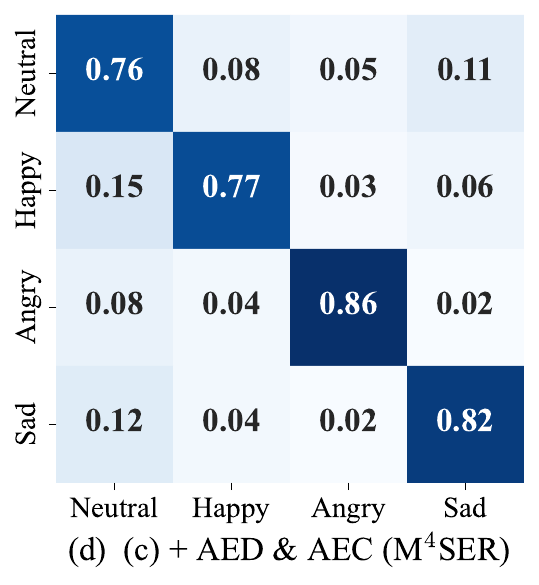}
\vspace{-3mm}
\caption{Confusion matrices obtained using IEMOCAP datasets. We utilize the results from five-fold cross-validation. Columns represent predicted labels and rows represent true labels.}
\label{fig:confusion_matrix}
\vspace{-5mm}
\end{figure*}
% =======confusion=======================

\textbf{Representation Analysis.}
We find that our M$^4$SER method performs better when using ASR text than when using GT text, as presented in Table \ref{tab:attn_analysis}. To investigate this finding, we visualize representation weights for two examples from IEMOCAP, as shown in Fig. \ref{fig:attention_visualization}. We display representations of each modality across the temporal level in two learning spaces, accumulating hidden features of $H_S$, $H_T$, $\hat{H}_S^{spe}$, and $H_T^{spe}$ into one dimension.
Fig. \ref{fig:attention_visualization}(a) illustrates that during single-modal learning with ASR text, representation weights in the text modality mainly focus on the word ``oh", whereas in the speech modality, representation weights concentrate towards the sentence's end. After cross-modal learning, we observed a significant decrease in ``oh" weight in the text modality. This change occurs as the model detects errors or inaccuracies in ASR transcription using AED and AEC modules, shifting representation weights to other words such as ``that's". Owing to limited emotional cues in the text, the model relies more on speech features, especially anger-related cues such as intonation and volume, critical for accurate anger recognition. Conversely, with GT text, after cross-modal learning, attention in the text modality focuses on words conveying positive emotions such as ``amusing", whereas speech modality distribution remains largely unchanged. Without ASR error signals, the model leans towards these textual features, leading to a bias towards happiness in emotion recognition.

Similarly, Fig. \ref{fig:attention_visualization}(b) shows inconsistency between GT text and speech features. GT text contains words (e.g., ``don't know") indicating confusion and uncertainty, whereas speech features exhibit significant emotional fluctuations. This mismatch complicates feature alignment in cross-modal learning, resulting in erroneous anger classification. In contrast, ASR text shows higher consistency with speech features, highlighting emotional words (``freaking out", ``what the hell", and ``all of a sudden") despite errors and simplifications. Representation weights confirm high consistency between ASR text and speech features across multiple regions, enhancing cross-modal learning accuracy and correct emotion identification as ``Happy" (``Excited"). We demonstrate with dynamic time warping (DTW) that ASR transcriptions are more consistent with speech than GT transcriptions, as shown in Table \ref{tab:dtw}.

In summary, compared with ASR text, GT text exhibits two primary differences:
\begin{enumerate}
    \item \textbf{Lack of ASR error signals:} GT text lacks ASR errors, missing additional cues beneficial for emotion recognition. This limitation possibly hinders capturing intense emotional signals when relying solely on GT text.
    \item \textbf{Differences in feature weight distribution:} the absence of ASR errors alters feature weight distribution in GT text modality compared with ASR text, potentially leading to inaccurate emotion classification.
\end{enumerate}

\textbf{Confusion Matrix Analysis.} To investigate the impact of different modules on class-wise prediction, we visualize the averaged confusion matrices over 5-folds on IEMOCAP in Fig.~\ref{fig:confusion_matrix}. Compared with the baseline, introducing MSR and MIR significantly enhances the prediction of ``Happy" class, which often suffer from ambiguous acoustic-textual alignment. 
By further incorporating GAN and LCL strategies, we observe overall improvements across most emotion classes, particularly in ``Neutral''. However, the performance on the ``Happy'' class slightly drops, potentially owing to the acoustic-textual heterogeneity introduced by merging ``Excited'' into ``Happy'', which possibly reduces the effectiveness of local contrastive learning in capturing class-specific discriminative features.
Finally, integrating AED and AEC modules yields the most accurate and balanced predictions overall. We observe further improvements in ``Neutral" and ``Sad” classes, while the performance of ``Happy” slightly recovers compared with the previous setting. These results highlight the effectiveness of multi-task learning in mitigating ASR-induced errors and enhancing emotional robustness across modalities.

\vspace{-3mm}
\subsection{Cross-corpus Generalization Ability}
To evaluate the performance of M$^4$SER in real-world environments, we conduct cross-corpus emotion recognition experiments, which simulate the practical scenario of domain shift between different datasets. Following the experimental settings of previous works~\cite{latif2022multitask, latif2025can, rehman2023speech}, we adopt a transfer evaluation strategy where one corpus is used entirely for training, and 30\% of the target corpus is reserved for validation for parameter tuning, while the remaining 70\% is used for final testing.
In addition, as the IEMOCAP dataset contains only four emotion classes (``Neutral", ``Happy", ``Angry", ``Sad"), we restrict MELD to the same set of overlapping emotions for a fair comparison and discard the rest (``Suprise", ``Fear", ``Disgust"). This ensures consistent label space across corpora during both training and evaluation.

Table~\ref{tab:zero-shot-comparison} presents the cross-corpus generalization results. Compared with the reimplemented SAMS and MF-AED-AEC baselines, our full model M$^4$SER achieves the best performance in both IE$\rightarrow$ME and ME$\rightarrow$IE directions. Specifically, M$^4$SER outperforms the strong multimodal baseline by 4.3\% in ACC and 3.1\% in W-F1 under the IE$\rightarrow$ME setting, and by 5.1\% WA and 4.8\% UA in the ME$\rightarrow$IE setting.

To further evaluate the adaptability of M$^4$SER under limited supervision, we conduct few-shot domain adaptation experiments by sampling 5, 10, 20, and 60 labeled instances per class from the target-domain validation set. The remaining validation data is still used for model selection, while the test set remains fixed across all settings. As shown in Table~\ref{tab:few-shot-comparison}, M$^4$SER consistently outperforms SAMS and MF-AED-AEC across all shot settings. For example, with only 5 labeled samples per class, M$^4$SER achieves ACC of 57.9\% in the IE$\rightarrow$ME setting, already surpassing MF-AED-AEC with 20 shots. As the number of target samples increases, our model scales well and reaches ACC of 62.1\% and W-F1 61.5\% at 60-shot. In the ME$\rightarrow$IE setting, M$^4$SER reaches UA of 75.9\%, demonstrating excellent cross-domain adaptability.

These results demonstrate that M$^4$SER not only generalizes robustly across corpora with minimal tuning, but also effectively adapts to new domains under few-shot conditions.

\begin{table}[t]
\centering
\caption{Analysis of cross-corpus generalization ability on a 4-class emotion classification task (\%). We reimplement the method indicated by $^{\circ}$ and obtain the corresponding result. \textit{IE $\rightarrow$ ME} indicates that the IEMOCAP dataset is used for training and MELD is used for testing. Conversely, \textit{ME $\rightarrow$ IE} means the model is trained on MELD and evaluated on IEMOCAP.}
\label{tab:zero-shot-comparison}
\begin{tabular}{lcccc}
\toprule
\multirow{2}{*}{\textbf{Methods}} & \multicolumn{2}{c}{\textbf{IE $\rightarrow$ ME}} & \multicolumn{2}{c}{\textbf{ME $\rightarrow$ IE}} \\
\cmidrule(lr){2-3} \cmidrule(lr){4-5}
& \textbf{ACC} & \textbf{W-F1} & \textbf{WA} & \textbf{UA} \\
\midrule
\midrule
SAMS$^{\circ}$ \cite{hou2023semantic}    & 17.0 & 11.6    & 36.3    & 33.9    \\
MF-AED-AEC$^{\circ}$ \cite{10446548} & 56.0 & 54.2 & 62.5 & 60.6 \\
\midrule
Baseline (HuBERT + BERT)                      & 53.0 & 52.4 & 58.9 & 57.4 \\
\hspace{1em}+ MSR \& MIR                   & 54.7 & 53.2 & 60.8 & 59.9 \\
\hspace{2em}+ GAN \& LCL  & 55.6 & 53.9 & 61.9 & 60.7 \\
\rowcolor[HTML]{EFEFEF} % 灰色
\textbf{\hspace{3em}+ AED \& AEC (M$^4$SER) } & \textbf{57.3} & \textbf{55.5} & \textbf{64.0} & \textbf{62.2} \\

\midrule
$\Delta_{\text{Baseline}}$     & 4.3$\uparrow$ & 3.1$\uparrow$ & 5.1$\uparrow$ & 4.8$\uparrow$ \\

\bottomrule
\end{tabular}
\vspace{-5mm}
\end{table}

\begin{table}[htbp]
\centering
\caption{Analysis of few-shot domain adaptation ability for cross-corpus 4-class emotion recognition (\%). We reimplement the method indicated by $^{\circ}$ and obtain the corresponding result. \textit{IE $\rightarrow$ ME} and \textit{ME $\rightarrow$ IE} denote training on IEMOCAP and MELD, respectively, with the other used for testing.}
\label{tab:few-shot-comparison}
\begin{tabular}{lccccc}
\toprule
\multirow{2}{*}{\textbf{Methods}} & \multirow{2}{*}{\textbf{\# Shots}} & \multicolumn{2}{c}{\textbf{IE $\rightarrow$ ME}} & \multicolumn{2}{c}{\textbf{ME $\rightarrow$ IE}} \\
\cmidrule(lr){3-4} \cmidrule(lr){5-6}
& & \textbf{ACC} & \textbf{W-F1} & \textbf{WA} & \textbf{UA} \\
\midrule
\midrule
\multirow{4}{*}{SAMS$^{\circ}$ \cite{hou2023semantic}} 
 % & 0  & 56.1 & 58.9 & 65.3 & 63.1 \\
 & 5  & 51.6 & 49.0 & 39.0 & 37.2 \\
 & 10 & 52.5 & 49.2 & 38.2 & 38.3 \\
 & 20 & 52.3 & 52.3 & 39.9 & 35.0 \\
 & 60 & 59.9 & 56.8 & 44.4 & 45.3 \\
\midrule
\multirow{4}{*}{MF-AED-AEC$^{\circ}$ \cite{10446548}} 
 & 5  & 56.3 & 54.8 & 64.4 & 63.3 \\
 & 10 & 57.8 & 55.7 & 65.8 & 65.6 \\
 & 20 & 57.5 & 56.7 & 67.6 & 66.6 \\
 & 60 & 60.8 & 60.3 & 72.5 & 73.7 \\
\midrule
\multirow{4}{*}{\textbf{M$^4$SER}} 
 % & 0  & \textbf{57.8} & \textbf{60.5} & \textbf{67.1} & \textbf{65.4} \\
 & 5  & \textbf{57.9} & \textbf{56.0} & \textbf{66.4} & \textbf{64.4} \\
 & 10 & \textbf{58.7} & \textbf{58.0} & \textbf{66.7} & \textbf{67.7} \\
 & 20 & \textbf{59.4} & \textbf{58.8} & \textbf{69.0} & \textbf{68.7} \\
 & 60 & \textbf{62.1} & \textbf{61.5} & \textbf{74.0} & \textbf{75.9} \\
\bottomrule
\end{tabular}
\vspace{0.0em}
\begin{flushleft}
\footnotesize
\end{flushleft}
\vspace{-9mm}
\end{table}

\vspace{-2mm}
\section{Conclusion}
% \vspace{-1mm}
\label{sec:conclusion}
In this paper, we propose M$^4$SER, a novel emotion recognition method that combines multimodal, multirepresentation, multitask, and multistrategy learning. M$^4$SER leverages an innovative multimodal fusion module to learn modality-specific and modality-invariant representations, capturing unique features of each modality and common features across modalities. We then introduce a modality discriminator to enhance modality diversity through adversarial learning. Additionally, we design two auxiliary tasks, AED and AEC, aimed at enhancing the semantic consistency within the text modality. Finally, we propose a label-based contrastive learning strategy to distinguish different emotional features. Results of experiments on the IEMOCAP and MELD datasets demonstrate that M$^4$SER surpasses previous baselines, proving its effectiveness. In the future, we plan to extend our approach to the visual modality and introduce disentangled representation learning to further enhance emotion recognition performance. We also plan to investigate whether AED and AEC modules remain necessary when using powerful LLM-based ASR systems.

\vspace{-3mm}
\bibliographystyle{IEEEtran}
\bibliography{IEEEtran}

\end{document}